\documentclass[a4paper, 11pt]{article}
\usepackage[margin=.75in]{geometry}
\usepackage{amsmath,graphicx,outlines,setspace,lineno}
\usepackage{natbib}
\usepackage{rotating}
\begin{document}
\noindent {\LARGE \textbf{Rats optimally accumulate and discount evidence in a dynamic environment}}\newline

\noindent {\Large Alex T Piet$^1$, Ahmed El Hady$^{1,3,*}$, Carlos D Brody$^{1,2,3,*}$}\newline
{\small $1$. Princeton Neuroscience Institute, Princeton University, Princeton, United States. \newline 
$2$. Department of Molecular Biology, Princeton University, Princeton, United States. \newline
$3$. Howard Hughes Medical Institute, Princeton University, Princeton, United States. }  \newline
*correspondence should be addressed to Ahmed El Hady (ahady@princeton.edu) or Carlos D Brody (brody@princeton.edu) \newline

\noindent {\Large \textbf{Abstract}}\newline
How choices are made within noisy environments is a central question in the neuroscience of decision making. Previous work has characterized temporal accumulation of evidence for decision-making in static environments. However, real-world decision-making involves environments with statistics that change over time. This requires discounting old evidence that may no longer inform the current state of the world. Here we designed a rat behavioral task with a dynamic environment, to probe whether rodents can optimally discount evidence by adapting the timescale over which they accumulate it.  Extending existing results about optimal inference in a dynamic environment, we show that the optimal timescale for evidence discounting depends on both the stimulus statistics and noise in sensory processing. We found that when both of these components were taken into account, rats accumulated and temporally discounted evidence almost optimally. Furthermore, we found that by changing the dynamics of the environment, experimenters could control the rats' accumulation timescale, switching them from accumulating over short timescales to accumulating over long timescales and back. The theoretical framework also makes quantitative predictions regarding the timing of changes of mind in the dynamic environment. This study establishes a quantitative behavioral framework to control and investigate neural mechanisms underlying the adaptive nature of evidence accumulation timescales and changes of mind. 
\newpage

\noindent {\Large \textbf{Introduction}}
\newline
Decision making refers to the cognitive and neural mechanisms underlying processes that generate choices. In our daily life, the processes of decision making are ubiquitous. Decision making has been a major focus in the neuroscience community because it bridges sensory, motor, and executive functions.  A well characterized decision making paradigm is that of ``evidence accumulation'' or ``evidence integration'' referring to the process by which the subject gradually processes evidence for or against different choices until making a well defined choice. Evidence accumulation is thought to underlie many different types of decisions from perceptual decisions \citep{brunton2013}, to social decisions  \citep{10.1371/journal.pcbi.1004371}, and to value based decisions \citep{Basten14122010}.  

Most behavioral studies to date have focused on evidence accumulation in stationary environments. In the case of stationary environments, the normative behavioral strategy used is perfect integration \citep{bogacz2006}, which refers to equal weighting of all incoming evidence across time. However, real world environments are complex and change over time. In this case, a strategy based on perfect integration will be suboptimal due to the changing statistics of the environment. Crucially, in a dynamic environment older observations may no longer reflect the current state of the world, and an observer needs to modify their inference processes to discount older evidence. Previous studies have demonstrated that humans can modify the timescales of evidence integration, adopting ``leaky'' integration when beneficial \citep{ossmy2013,glaze2015}. This observation opens many questions related to why and how subjects might alter their integration timescales. To answer ``why'' or normative questions, one would ideally like to develop a model that can be directly compared to the standard evidence accumulation models used in the decision making literature. Two recent studies have developed this connection to drift-diffusion models, and examined evidence accumulation in dynamic environments either in humans \citep{glaze2015,goldreview2017} or in ideal observer models \citep{kilpatrick2015}. Animal models of behavior facilitate investigation of ``how'' or mechanistic questions, by allowing measurement and perturbation of neural circuits. Here, we demonstrate that rats are capable of adopting the optimal integration timescale predicted by the recently developed modeling framework \citep{kilpatrick2015}, and we furthermore show that they can dynamically modulate their integration timescale according to changing environmental statistics. 

In the present study, we extend a previously published pulse-based accumulation of evidence task the ``Poisson clicks task'' \citep{brunton2013,erlich2014, hanks2015} to a dynamic environment. We refer to our task as the ``Dynamic clicks task''.  We extend results from the literature \citep{kilpatrick2015} to develop the optimal inference process for our task. The ideal observer is closely related to the ``drift-diffusion model'' used widely in the decision making literature \citep{bogacz2006,ratcliff2008}. The primary difference is that, in addition to integrating sensory evidence, the ideal observer discounts accumulated evidence at a rate proportional to the volatility of the environment, and the reliability of each evidence pulse. The reliability of each pulse is determined by the stimulus statistics (e.g., the pulse rates), as well as noise in the subject's sensory transduction process. While the exact origin of sensory noise is unclear, quantitative modeling can separate sensory noise from other types of noise \citep{brunton2013}. Here, we use sensory noise to refer to noise that scales with the amount of evidence. The role of sensory noise in decision making processes is a relatively unexplored area. Studies in the literature are beginning to document under what circumstances subjects modify their behavior based on noise in the sensory evidence \citep{GURECKIS2009180,zylerberg2016}.  

Using high-throughput behavioral training, we trained rats to perform this task. With a combination of quantitative methods, we find that rats' adaptation to the dynamic environment is such that they adopt the optimal timescale for evidence accumulation. Our findings establish rats as an adequate animal model for evidence accumulation in a dynamic environment. Training rodents on state of the art cognitive tasks opens up the opportunity to understand the neuronal mechanisms underlying complex behavior. Rodents can be trained in a high throughput manner, are amenable to genetic manipulation, are accessible to electrophysiological and optogenetic manipulations, and a large number of experimental subjects can be used. Finally, the dynamic clicks task opens up the opportunity to study the neural underpinnings of evidence integration in a dynamic environment as this task gives the experimentalist a unique quantitative handle over the integration timescale of the animals.  \newline

\begin{figure}
\centering
\includegraphics[width=.6\textwidth]{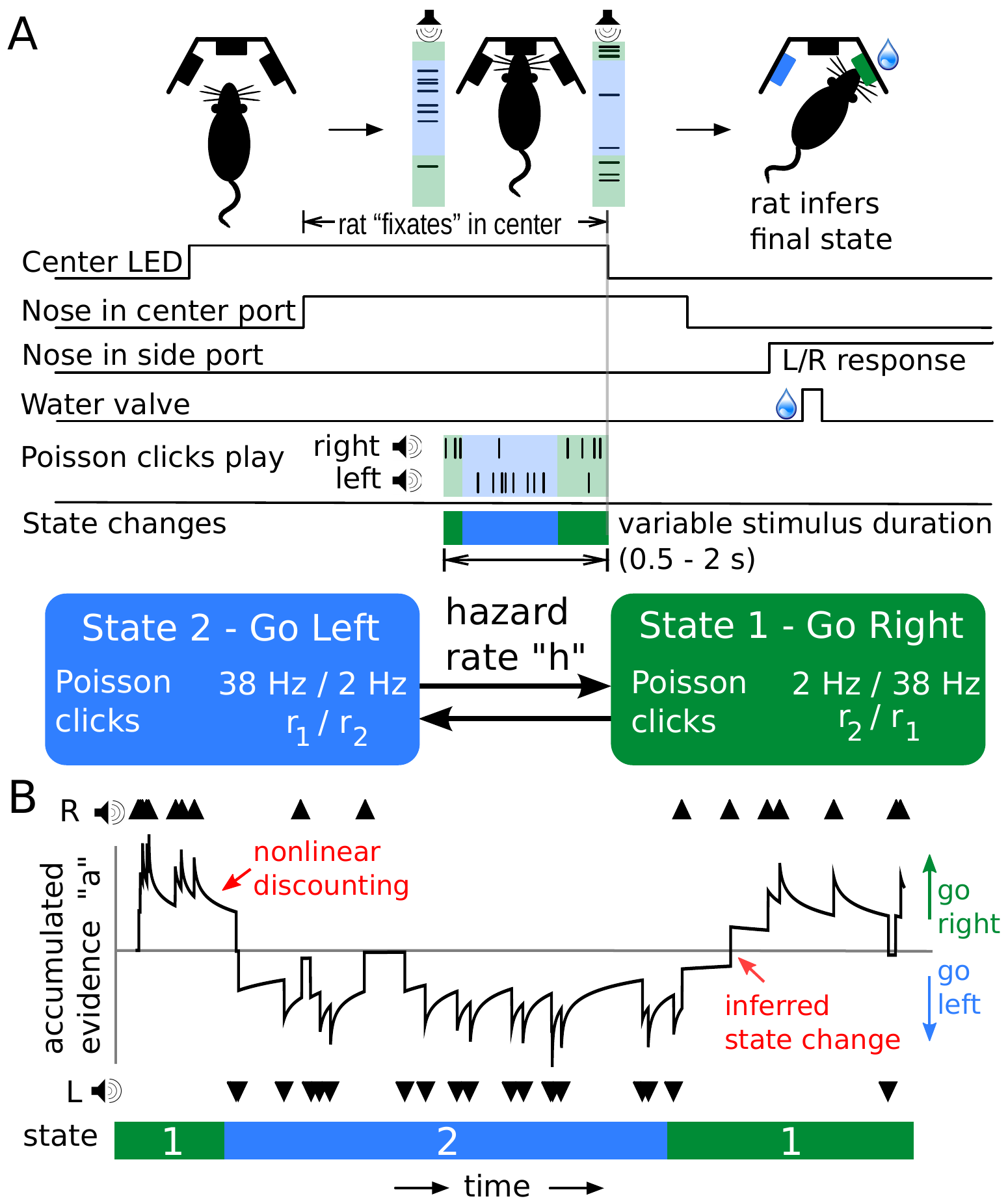}
\caption{\textbf{Dynamic Clicks Task structure and example trial.} (A) Schematic of task events and timing. A center light illuminates indicating the rat may initiate a trial by poking its nose into a center port. Auditory clicks are generated from state-dependent Poisson processes (the two states are schematized by light green and light blue backgrounds) and played concurrently from left and right speakers. The hidden state toggles between two states according to a telegraph process with hazard rate $h$. When the auditory clicks end, and the center light turns off, the rats must infer which of the two states the trial ended in and report their decision by poking into one of two reward ports. Trials have random durations so the rat must be prepared to answer at all time points. (B) An example trial illustrates features of the task. The hidden state transitions randomly, and the auditory clicks are generated accordingly. The optimal inference process (black line; see text for its derivation) accumulates clicks, and discounts accumulated evidence proportionally to the volatility of the environment and click statistics. For the optimal process, a choice is generated at the end of the trial according to whether the optimal inference variable is above or below 0.}
\end{figure}

\noindent {\Large \textbf{Results}}\newline
\noindent{\textbf{A dynamic decision making task}} \newline
We developed a decision making task that requires accumulating noisy evidence in order to infer a state that is hidden, and dynamic.  Rats were trained to  infer, at any moment during the course of a trial, which of two states the environment was in at that moment. These could be either a state in which randomly-timed auditory clicks were played from a left-speaker at a high rate and right speaker clicks were played at a low rate, or its inverse (low rate on the left, high rate on the right).  In more detail, in each trial of our task, we first illuminate a center light inside an automated operant chamber, to indicate that the rat may start the trial by nose-poking into the center port. Once the rat enters the center port, auditory clicks play from speakers positioned on the left and right sides of the rat. The auditory clicks are generated from independent Poisson processes. Importantly, the left and right side Poisson rate parameters are dependent on a hidden state that changes dynamically during the course of each trial. This is in sharp contrast to previous studies where the Poisson click rates are constant for the duration of each trial \citep{brunton2013,erlich2014,hanks2015}. Within each trial, the dynamic environment is in one of two hidden states $S^1$, and $S^2$, each of which has an associated left and right click generation rate ($S^1$: rates $r^1_L$ and $r^1_R$, respectively; $S^2$: rates $r^2_L$ and $r^2_R$). In this study  $S^1$ and $S^2$ were symmetric  ($r_R^1 = r_L^2 =$ high rate $r_1$ and $r_R^2 = r_L^1 = $ low rate $r_2$). Each trial starts with equal probability in one of the two states, and switches stochastically between them at a fixed ``hazard rate'' $h$. On each time step, the switch probability is given by $h\Delta t$,  (with $\Delta t$ kept small enough that  $h\Delta t << 1$). At the end of the stimulus period, the auditory clicks end, and the center light turns off, indicating the rat must make a left or right choice by entering one of the side reward ports. The rat is rewarded with a water drop for correctly inferring the hidden state at the end of the stimulus period (if $S^1$, go right; if $S^2$, go left).  The stimulus period duration is variable on each trial ($0.5 - 2$ seconds), so the rat must be prepared to infer the current hidden state at all times. Figure 1 shows a schematic of task events, as well as an example trial. Rats trained every day, performing 150-1000 self-paced trials per day.\newline

\noindent{\textbf{Optimal inference in a dynamic environment}} \newline
Here we derive the optimal procedure for inferring the hidden state. Optimality, in this setting, refers to reward maximizing. 
Given that each trial's duration is imposed by the experimenter and thus fixed to the rat, maximizing reward is equivalent to maximizing accuracy \citep{bogacz2006}. We build on results from \citealt{kilpatrick2015}, but a basic outline is repeated here for continuity. Mathematical details can be found in the supplementary materials. 

Before diving into the derivation, it is worth building some intuition. Because the hidden state is dynamic, auditory clicks heard at the start of the trial are unlikely to be informative of the current state. However, because state transitions are hidden, an observer doesn't know how far back in time observations are still informative of the current state. Our derivation derives the optimal weighting of older evidence. We first consider observations in discrete timesteps of short duration $\Delta t$. Within each timestep, a momentary evidence sample $\epsilon$ is generated. This sample is either a click on the left, a click on the right, no clicks, or a click on both sides (we will consider $\Delta t$ small enough that $r_1\Delta t << 1$ and $r_2 \Delta t << 1$ so that multiple clicks are not generated within one timestep). 

Following \citealt {kilpatrick2015}, the probability of being in State 1 at time $t$, given all observed samples up to time $t$:\begin{eqnarray}\label{state_likelihood}
P\left(S^1| \epsilon_{1 \ldots t} \right)  \propto P(\epsilon_t|S^1)\left(\left(1-h\Delta t \right)P\left(S^1| \epsilon_{1 \ldots t-1} \right) + h\Delta tP\left(S^2| \epsilon_{1 \ldots t-1} \right)   \right).
\end{eqnarray} 
We can interpret this equation as the probability of being in State 1 given all observed evidence up to time t ($P\left(S^1| \epsilon_{1 \ldots t} \right)$) is proportional to the probability of observing the evidence sample at time $t$ given State 1 ($P(\epsilon_t|S^1)$) times the independent probability that we were in State 1 given evidence from timesteps $1 \ldots t-1$ ($P(S^1|\epsilon_{1 \ldots t-1})$). This second term is decomposed into two terms which depend on the probability of remaining in the same state from the last time step ($\left(1-h\Delta t \right)P\left(S^1| \epsilon_{1 \ldots t-1} \right)$) and the probability of changing states after the last time step ($h\Delta tP\left(S^2| \epsilon_{1 \ldots t-1} \right)$).

Combining the probability of each state into a ratio, we can write the posterior probability ratio ($R_t$) of the current state given all previous evidence samples $\epsilon_{1 \ldots t}$:
\begin{eqnarray}\label{ratio}
R_t &=&\frac{P(S^1 | \epsilon_{1 \ldots t})}{P(S^2| \epsilon_{1 \ldots t})} = \frac{P(\epsilon_t | S^1)}{P(\epsilon_t | S^2)}\left(\frac{\left(1-h\Delta t \right)R_{t-1} + h\Delta t  }{\left(h\Delta t\right) R_{t-1} + 1 -h \Delta t  }\right).
\end{eqnarray}
Observe that in a static environment ($h = 0$), the term on the far right simplifies to $R_{t-1}$ and \eqref{ratio} becomes the statistical test known as the Sequential Probability Ratio Test (SPRT) \citep{wald1945,barnard1946, bogacz2006}. A recent study demonstrated that monkeys could accurately perform a literal instantiation of the SPRT \citep{kira2015}. When $h \neq 0$, the more complicated expression reflects the fact that previous evidence samples might no longer be informative of the current state, in a manner proportional to the environmental volatility $h$. 

In order to compare \eqref{ratio} to standard decision making models like the drift-diffusion model (DDM) we will transform the expression into a differential equation.  We can accomplish this by taking the logarithm of \eqref{ratio}, then  substituting $\hat{a} = \log\left(R\right)$, and finally taking the limit of $\Delta t$ goes to $0$ (See \citealt{kilpatrick2015} and supplementary materials for details):
\begin{eqnarray}\label{log_ratio}
d\hat{a} &=& \log\left(\frac{P(\epsilon_t | S^1)}{P(\epsilon_t | S^2)}\right) - 2h\sinh\left(\hat{a}\right)dt.
\end{eqnarray} 
This differential equation describes the evolution of the log-probability ratio of being in each of the two hidden states $\left(\hat{a} = \log\left(\frac{P(S^1 | \epsilon_{1 \ldots t})}{P(S^2| \epsilon_{1 \ldots t})} \right)\right)$: $\hat{a} > 0$ indicates more evidence for $S^1$, while $\hat{a} < 0$ indicates more evidence for $S^2$. Momentary evidence samples $\epsilon_t$ are incorporated into the log-probability ratio through the evidence term ($\log\left(\frac{P(\epsilon_t | S^1)}{P(\epsilon_t | S^2)}\right)$). The previously accumulated evidence is forgotten by a nonlinear discounting term ($- 2h\sinh\left(\hat{a}\right)$) (See Fig 2C). The evidence discounting reduces the effect of older evidence, weighting recent evidence more. This discounting reflects the fact that older evidence may no longer be informative of the current state of the environment. 
In a static environment ($h=0$), the discounting term is eliminated, and the ideal observer perfectly integrates the momentary evidence samples. In analysis of the static decision making models, the evidence term is commonly approximated by its expectation (drift) and variance (diffusion), transforming \eqref{log_ratio} into the Drift-Diffusion Model (DDM) for decision making \citep{bogacz2006}. 

From this point on our derivation departs from existing results in the literature. In order to develop a deeper understanding of the optimal inference on our task, we will evaluate the evidence term. Because of the discrete nature of the Poisson evidence, this term can be precisely evaluated for each evidence sample  in a way that is not possible in other decision making tasks. In a small sample window of duration $\Delta t$, the probability of a Poisson event is $r\Delta t$, where $r$ is the parameter of the Poisson process (provided $r\Delta t << 1$). In our task a momentary sample $\epsilon_t$ is the result of two independent Poisson processes and can take on four possible values: a click on both sides, a click on the right, a click on the left, or no clicks. Evaluating the evidence term for these four conditions:

\noindent A click on both sides
\begin{align}
\log\frac{P(\epsilon_t | S^1)}{P(\epsilon_t | S^2)} &= \log\frac{P\left(\text{click-R}|S^1\right)P\left(\text{click-L}|S^1\right)}{P\left(\text{click-R}|S^2\right)P\left(\text{click-L}| S^2\right)}= \log\frac{\left(r_1\Delta t\right)\left( r_2\Delta t\right)}{\left(r_2 \Delta t\right)\left(r_1\Delta t\right)} = 0.\\ \intertext{No clicks}
\log\frac{P(\epsilon_t | S^1)}{P(\epsilon_t | S^2)} &= \log\frac{P\left(\text{no-click-R}|S^1\right)P\left(\text{no-click-L}|S^1\right)}{P\left(\text{no-click-R}|S^2\right)P\left(\text{no-click-L}| S^2\right)}= \log\frac{(1-r_1\Delta t)(1- r_2\Delta t)}{(1-r_2\Delta t)(1- r_1\Delta t)} = 0.\\ \intertext{A click on the right}
\log\frac{P(\epsilon_t | S^1)}{P(\epsilon_t | S^2)} &= \log\frac{P\left(\text{click-R}|S^1\right)P\left(\text{no-click-L}|S^1\right)}{P\left(\text{click-R}|S^2\right)P\left(\text{no-click-L}| S^2\right)} = \log\frac{(r_1\Delta t)(1- r_2\Delta t)}{(r_2\Delta t)(1- r_1\Delta t)} \equiv + \kappa\left(r_1, r_2\right). \\ \intertext{A click on the left}
\log\frac{P(\epsilon_t | S^1)}{P(\epsilon_t | S^2)} &= \log\frac{P\left(\text{no-click-R}|S^1\right)P\left(\text{click-L}|S^1\right)}{P\left(\text{no-click-R}|S^2\right)P\left(\text{click-L}| S^2\right)} = \log\frac{(1-r_1\Delta t)(r_2\Delta t)}{(1-r_2\Delta t)(r_1\Delta t)} \equiv - \kappa\left(r_1, r_2\right) .
\end{align}
We define the function $\kappa(r_1,r_2)$ to be the increase in the log-probability ratio from the arrival of a single click on the right, given click rates $r_1$, $r_2$. The function $\kappa$ tells us how reliably each click indicates the hidden state. This is easily seen when letting $\Delta t \rightarrow 0$, so $\kappa \rightarrow \log\frac{r_1}{r_2}$. If the click rates $r_1$ and $r_2$ are very similar (so $\kappa$ is small) then we expect many distractor clicks (clicks from the smaller click rate that do not indicate the correct state), so an individual click tells us little about the underlying state. On the other hand, if the click rates are very different (so $\kappa$ is large) then we expect very few distractor clicks, so an individual click very reliably informs the current state. In the limit of one of the click rates going to zero: $\kappa \rightarrow \infty$, and a single click tells us the current state with absolute certainty. In our task, the two click rates $r_1$ and $r_2$ always sum to $40$ hz.  Figure 2A shows $\kappa$ as a function of the click rates. 

Re-writing the log-evidence term in \eqref{log_ratio} in terms of $\kappa$ and using $\delta_{L/R,t}$ to represent the left/right click times, we can summarize across all four conditions:
\begin{eqnarray} \label{opt_accum_no_scale}
d\hat{a} &=& \kappa\left(r_1,r_2 \right)\left(\delta_{R,t} - \delta_{L,t}\right)  - 2h\sinh(\hat{a})dt.
\end{eqnarray}
We can then rescale equation \eqref{opt_accum_no_scale} by $\kappa$,  let $a=\frac{\log(R)}{\kappa}$, to put our evidence accumulation equation in units of clicks:
\begin{eqnarray} \label{opt_accum}
da &=& \delta_{R,t} - \delta_{L,t}  - \frac{2h}{\kappa}\sinh(\kappa a)dt.
\end{eqnarray} Here $\delta_{L/R,t}$ are trains of delta functions at the times of the left and right clicks. Equation \eqref{opt_accum} has a simple interpretation, sensory clicks are integrated ($\delta_{R,t} - \delta_{L,t}$), while accumulated evidence is discounted (-$\frac{2h}{\kappa}\sinh(\kappa a)$) proportionally to the volatility of the environment $(h)$, and the reliability of each click $(\kappa)$. This interpretation also allows for a simple assay of behavior: do rats adopt the optimal discounting timescale? We will present two quantitative methods for measuring the rats discounting timescales. However, before examining rat behavior, we need to examine the impact of sensory noise on optimal behavior. \newline

\begin{figure}
\includegraphics[width=.9\textwidth]{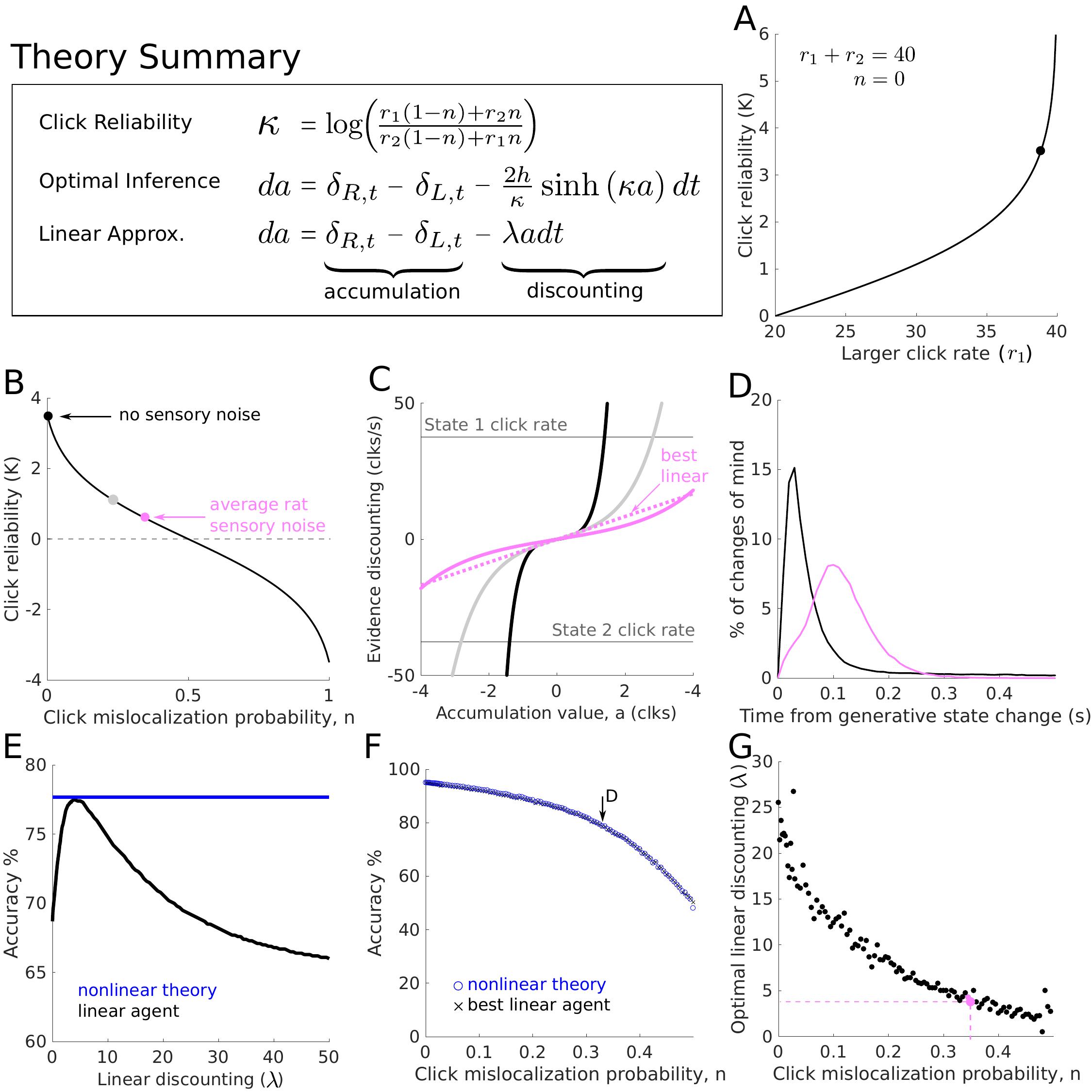}
\caption{\textbf{Optimal discounting rates depends on click reliability and can be well-approximated by linear discounting} (A) The reliability $\kappa$ of each click depends on the Poisson click rates $r_1$ and $r_2$. If the click rates are very similar, each click is not very informative about the underlying state. Black dot shows the rates used in the study. (B) The reliability of each click also depends on how consistently each click can be correctly localized to the side that generated it. At 50\% mislocalization each click contains no information about the current state, so $\kappa =0$. The light pink dot is the average level of sensory noise reported in Brunton 2013. The grey dot is half of the sensory noise in Brunton 2013. (C) Discounting functions for the three sensory noise levels in B (same colors). Increasing sensory noise causes the discounting functions to weaken. Horizontal lines show average clks/sec in each of the two states. (D) Histogram of changes of mind produced by the optimal inference equation. Timing is relative to the last change in the hidden state. (Black) Inference without sensory noise, (pink) inference with average rat level of sensory noise. (E) The optimal nonlinear discounting function can be approximated by a linear discounting function. If the linear discounting function is tuned appropriately, accuracy is close to the full nonlinear function. (F) Comparison between optimal nonlinear discounting function (blue) and the best linear approximation (black), in terms of average accuracy for different noise levels. The best linear approximation is effectively equivalent. Arrow indicates parameter values used in panel D.  (G) The best linear discounting rate $\lambda$ as a function of sensory noise. Increasing sensory noise decreases the discounting rate. The best linear function is found numerically on a set of 30k trials, which produces some variability for different noise levels. Pink dot indicates average rat sensory noise. }
\end{figure}

\noindent{\textbf{Sensory noise decreases click reliability}} \newline
The function $\kappa\left(r_1,r_2\right)$ tells us how reliably each click indicates the underlying state as a function of the click generation rates $r_1$ and $r_2$. The computation above of $\kappa$ assumes that each click is detected and correctly localized as either a left or right click with perfect accuracy. Previous studies using pulse-based evidence demonstrate that rats have significant sensory noise \citep{brunton2013,scott2015}. The term sensory noise in the context of these studies refers to sources of errors that scale with the number of pulses of evidence. Sensory noise was measured by fitting parametric models that included a parameter for how much uncertainty in the accumulation variable was increased due to each pulse of evidence. The exact biological origin of this noise remains unclear. It could arise from sensory processing errors, or from disruption of coding in the putative integration circuit at the moment of pulse arrival. Regardless of its origins, sensory noise is a significant component of rodent behavior.

We will now show that sensory noise decreases how reliably each click indicates the underlying state. 
While sensory noise can be modeled in many ways, primarily the mislocalization of clicks changes the click reliability. We analyze the cases of Gaussian noise on the click amplitudes and missing clicks, and provide a general argument for mislocalization in the supplementary materials. Mislocalization refers to how often clicks are incorrectly localized to the other speaker (hearing a click from the left and assigning it to the right). For intuition, consider that if a rat could never tell whether a click was played from the right or left then each click would never indicate any information about the underlying state.  We can again evaluate the log-evidence term, this time including the probability of click mislocalization ($n$):

\noindent{A click on the right}
\begin{align}
\log\frac{(r_1\Delta t)(1-n)(1- r_2\Delta t)+(1-r_1\Delta t)(r_2\Delta t)(n)  }{(r_2\Delta t)(1-n)(1- r_1\Delta t)+(1-r_2\Delta t)(r_1\Delta t)(n)} &= + \kappa\left(r_1, r_2,n\right). \\ \intertext{A click on the left}
\log\frac{(r_2\Delta t)(1-n)(1- r_1\Delta t)+(1-r_2\Delta t)(r_1\Delta t)(n)}{(r_1\Delta t)(1-n)(1- r_2\Delta t)+(1-r_1\Delta t)(r_2\Delta t)(n)  }&= - \kappa\left(r_1, r_2,n\right). 
\end{align} The terms for no clicks, or clicks on both sides evaluate to 0. As in the case with no sensory noise, the log-evidence is either 0, or has value $\kappa$. We can simplify the expression for $\kappa$ by letting $\Delta t \rightarrow 0$:
\begin{eqnarray}
\kappa(r_1,r_2,n) = \log\frac{r_1(1-n)+r_2n}{r_2(1-n)+r_1n}.
\end{eqnarray}
Sensory noise decreases how reliably each click informs the underlying state in the trial, increasing $n$ decreases $\kappa$. If $n=0$, we recover the original $\kappa$ derived without noise. If $n=0.5$, then each click is essential heard on a random side, and therefore contains no information so $kappa = 0$. If $n = 1$, then we simply flip the sign of all clicks. 

Previous studies using the same auditory clicks have shown that rats have significant sensory noise. Figure 2B shows $\kappa$ against $n$, and highlights the average sensory noise, and corresponding $\kappa$, found in a previous study \citep{brunton2013}. \newline

\noindent{\textbf{Lower click reliability requires longer integration timescales}} \newline
The discounting term of equation \eqref{opt_accum} has $\kappa$ in the denominator as well as the argument of the $\sinh$ term. As a result, it is not clear how decreasing the click reliability $\kappa$ changes the behavior of the optimal inference agent. To gain insight, consider that if evidence is very reliable, accurate decisions can be made by only using a few clicks from a small time window. However, if evidence is unreliable, a longer time window must be used to average out unreliable clicks. This intuition is confirmed by plotting the discounting function for a variety of evidence reliability values (Figure 2C). Decreasing reliability weakens the evidence discounting term creating longer integration timescales. See the supplementary materials for more details. \newline

\noindent{\textbf{Evidence discounting leads to changes of mind}} \newline
 The optimal inference equation attempts to predict the hidden state. As the hidden state dynamically transitions, we expect the inference process to track, albeit imperfectly, the dynamic transitions. From the perspective of a subject this dynamic tracking leads to changes of mind in the upcoming choice. Through the optimal inference process we can predict the timing of changes of mind by looking for times when the sign of the inference process changes ($\text{sign}(a)$). The presence of sensory noise slows the integration timescale, and thus slows the timing of changes of mind. Figure 2D shows the predicting timing of changes of mind with and without sensory noise.\newline

\noindent{\textbf{Linear approximation to nonlinear discounting function is very accurate }} \newline
The full nonlinear discounting function ($-\frac{2h}{\kappa}\sinh\left(\kappa a\right)$), is complicated. In order to aid our analysis of rat behavior, we will consider a linear approximation to the discounting function ($-\lambda a$), where $\lambda$ gives the discounting rate. There are many possible linear approximations with different slopes. A linear approximation using the slope of $\sinh$ at the origin will fail to capture the strong discounting farther from the origin. We found the best linear approximation numerically. 

Figure 2E shows, for a particular noise level and click rates, the accuracy of a range of linear discounting agents against the full nonlinear agent. If $\lambda$ is tuned correctly, the linear agent accuracy is very close to the full nonlinear function. We find this to be true across a wide range of noise values (Figure 2F). While the optimal linear strength at each noise level changes (Figure 2G), the accuracy is always very close to that of the full nonlinear theory. It is important to note that a linear approximation in general will not always be close in accuracy to the full nonlinear theory, but for our specific click rate parameters it is an accurate approximation. See \citealt{kilpatrick2015} for examples of evidence statistics for which the linear approximation does not fit as well. 

Given that a linear discounting function matches the accuracy of the full nonlinear model, we will analyze rat evidence discounting behavior by looking for the appropriate discounting rate or equivalently the appropriate integration timescale. Specifically, we will compare the rat behavior to this linear discounting equation: 
\begin{eqnarray} \label{opt_linear_accum}
da &=& \delta_{R,t} - \delta_{L,t}  - \lambda a dt,
\end{eqnarray}
where $\lambda$ is the discounting rate and $\frac{1}{\lambda}$ is the integration timescale. We did not examine whether rats demonstrate nonlinear evidence discounting because the linear approximation in our task is effectively indistinguishable from the full nonlinear theory.\newline

\begin{figure}
\centering
\includegraphics[width=.75\textwidth]{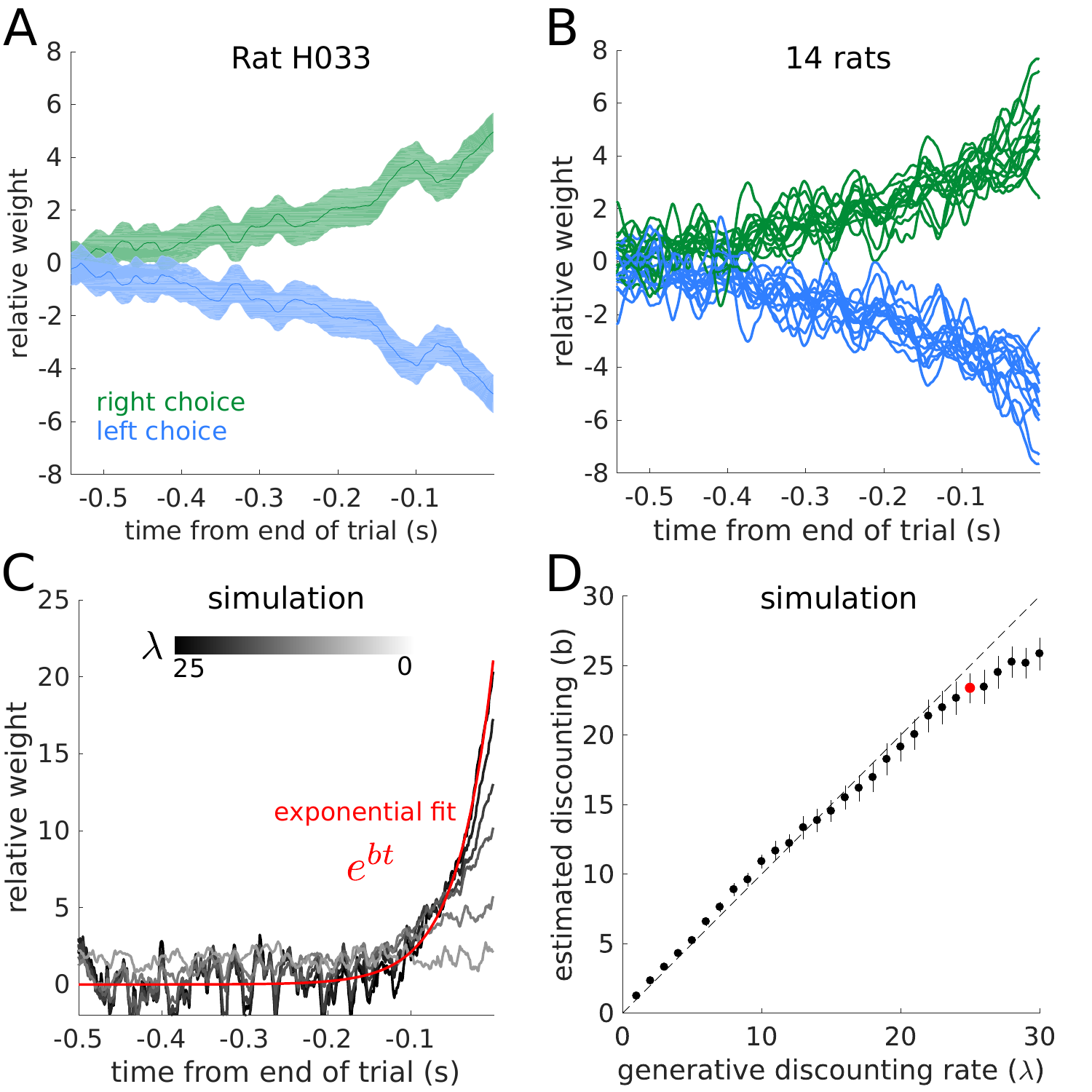}
\caption{\textbf{Rats discount evidence.} 
(A) Reverse correlation curves for an example rat reveals how clicks at each time point influence the rat's decision. (B) Reverse correlation curves for 14 rats. Error bars are omitted for clarity. (C) Reverse correlation curves for a range of simulated linear discounting agents. Black to white lines indicate increasing discounting rates ($\lambda$). Only the reverse correlation curve for the right choice are shown for clarity. Each curve was fit with an exponential function (example red). The fit parameters are used in part D. (D) Exponential fit to each discounting agent recovers the generative linear discounting rate. Example in part C show with red dot.}
\end{figure}

\noindent{\textbf{Psychophysical reverse correlation reveals the integration timescale}} \newline
Psychophysical reverse correlation is a commonly used statistical method to find what aspects of a behavioral stimulus influence a subject's choice. Here we use reverse correlation to find the integration timescale used by the rats. We then normalized the reverse correlation curve to have an area under the curve equal to one. This step lets the curves be interpreted in units of effective weight at each time point. A flat reverse correlation curve indicates even weighting of evidence across all time points. Previous studies in a static environment find rats with flat reverse correlation curves \citep{brunton2013,hanks2015,erlich2014}. Figure 3A shows the reverse correlation for an example rat in a dynamic environment. The stimulus earlier in the trial is weighted less than the stimulus at the end of the trial indicating evidence discounting. Figure 3B shows the mean reverse correlations for all rats in the study. Figure 3C shows the reverse correlation curves from a family of linear discounting agents $\left(da = \delta_R - \delta_L -\lambda a dt\right)$, with $\lambda$ ranging from 0 to 30. The curves were generated from a synthetic dataset of 20,000 trials. The weaker the discounting rate, the flatter the reverse correlation curves. To quantify the discounting timescale from the reverse correlation curves, an exponential function $e^{bt}$ was fit to each curve. The parameter $b$ reliably recovers the discounting rate $\lambda$ (Figure 3D). \newline

\begin{figure}
\centering
\includegraphics[width=.75\textwidth]{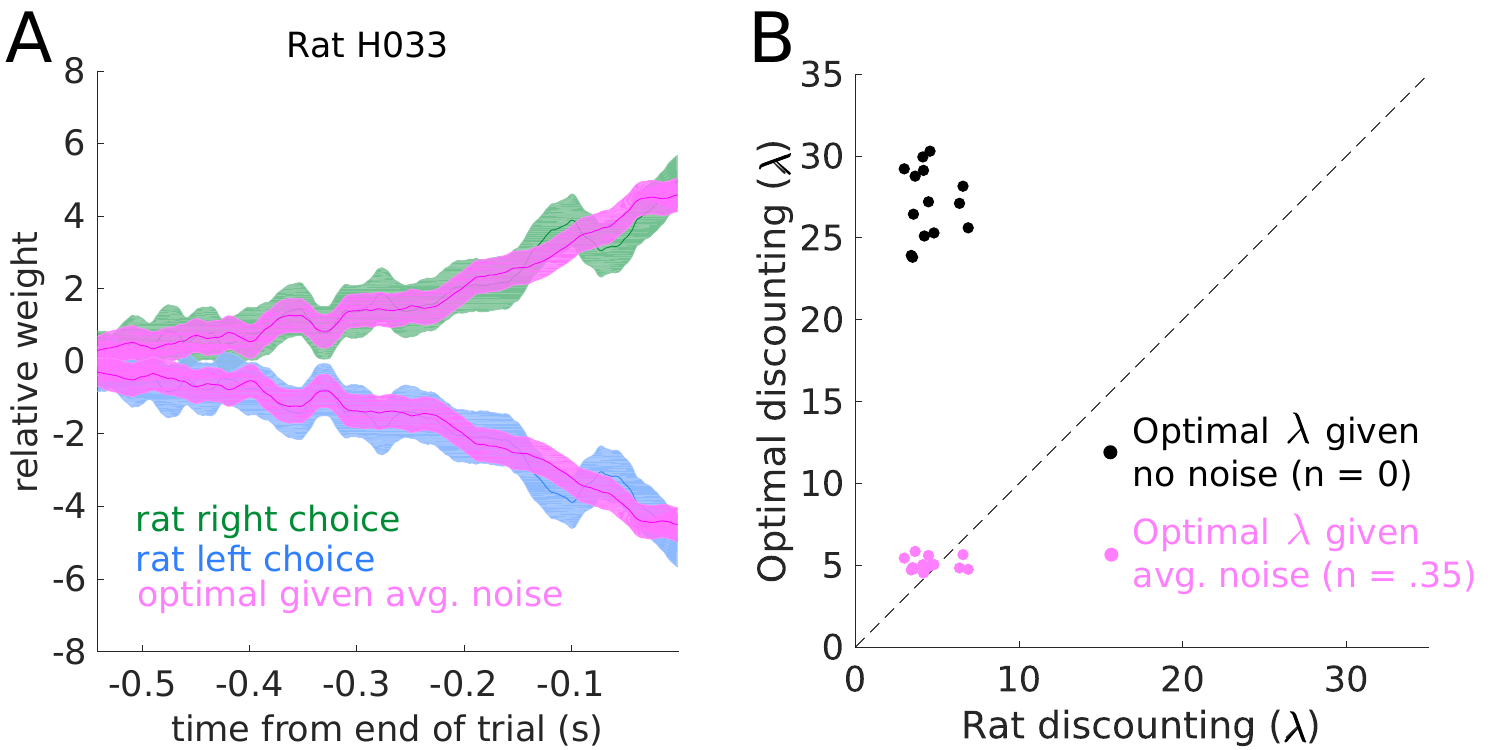}
\caption{\textbf{Rats optimally discount evidence.} (A) Example reverse correlation curve for one rat, and the reverse correlation curve from the optimal inference agent with the average rat sensory noise. The optimal inference agent was simulated on the same trials the rat performed. (B) Quantification of discounting timescales. When factoring average sensory noise, the rats adopt the optimal timescale. The variability in optimal discounting rates is a result of measuring the reverse correlation curves on a different set of trials each rat actually performed. }
\end{figure}

\noindent{\textbf{Rats adapt to the optimal timescale}} \newline
To compare each rat's evidence discounting timescale to the optimal inference equation, we simulated the optimal inference agent on the trials each rat experienced. We then computed the reverse correlation curves for both the rats and the optimal agent (Figure 4A).  
To quantitatively compare timescales, we then fit an exponential function to each of the reverse correlation curves. Rat behavior was compared with two optimal agents. The first optimal agent assumes no sensory noise; while the second agent uses the optimal timescale given the average level of sensory noise across rats reported in \citealt{brunton2013} (Figure 4B).  When the average level of sensory noise is taken into account, the rats match the optimal timescale. The reverse correlation analysis shows that rats are close to optimal given the average level of sensory noise in a separate cohort of rats. \newline

\noindent{\textbf{A quantitative behavioral model captures rat behavior}} \newline  
In order to extend our analysis to examine individual variations in noise level and integration timescales, we fit a behavioral accumulation of evidence model from the literature to each rat \citep{brunton2013,hanks2015,erlich2014}. This model generates a moment-by-moment estimate of a latent accumulation variable. The dynamical equations for the model are given by:
\begin{eqnarray} \label{bing_model}
da &=& \left(\delta_{R,t}\cdot \eta_R\cdot C - \delta_{L,t}\cdot \eta_L\cdot C\right)dt - \lambda a dt + \sigma_a dW,\\
\frac{dC}{dt} &=& \frac{1-C}{\tau_{\phi}} + \left(\phi -1\right)C\left(\delta_{R,t} + \delta{L,t}\right).
\end{eqnarray}
At each moment in a trial, the model generates a distribution of possible accumulation values $P(a | t, \delta_R, \delta_L)$. In addition to the click integration and linear discounting that was present in our normative theory, this model also parameterizes many possible sources of noise. Each click has multiplicative Gaussian sensory noise, $\eta_{L/R} = \mathcal{N}\left(1,\sigma_s^2 \right)$. In addition to the sensory noise, each click is also filtered through an adaptation process, $C$. The adaptation process is parameterized by the adaptation strength $\phi$, and a adaptation time constant $\tau_{\phi}$. If $\phi > 1$ the model has facilitation of sequential clicks, and if $\phi < 1$ the model has depression of sequential clicks. The accumulation variable $a$ also undergoes constant additive Gaussian noise $\sigma_a$. Finally, the initial distribution of $a$ has some initial variance given by $\sigma_i$. See \citealt{brunton2013} for details on the development and evaluation of this model. One major modification to the model from previous studies is the removal of the sticky bounds $B$, which are especially detrimental to subject performance given the dynamic nature of the task. This model is a powerful tool for the description of behavior on this task because of its flexibility at characterizing many different behavioral strategies \citep{brunton2013,hanks2015,erlich2014}.

\begin{figure}
\centering
\includegraphics[width=.7\textwidth]{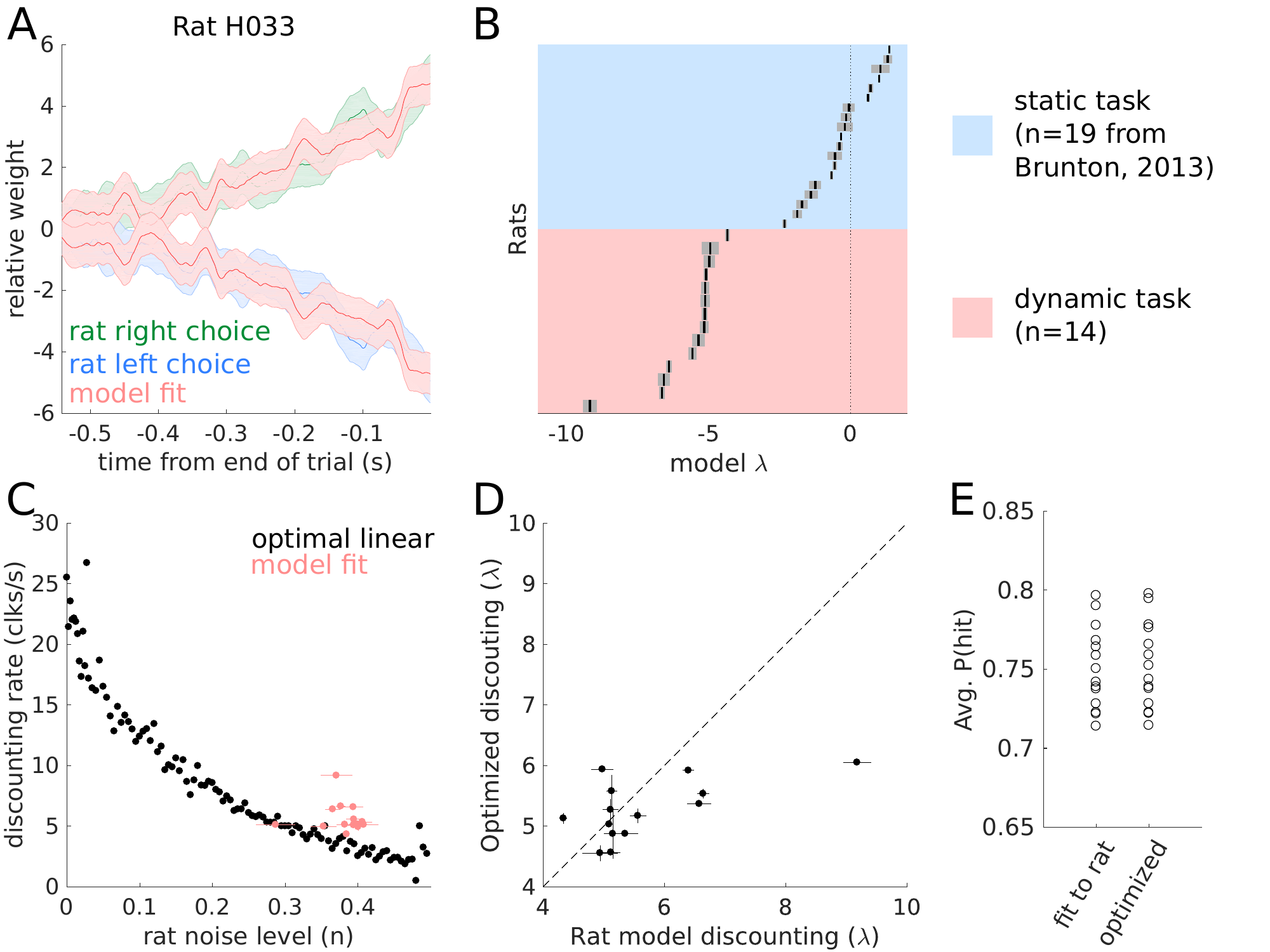}
\caption{\textbf{Quantitative model captures rat behavior, and shows optimal discounting} (A) Example reverse correlation curves generated by the quantitative model compared with a rat's behavior. (B) Best fitting discounting rates for rats trained on the dynamic task (orange), and for rats trained in a static environment (blue, data and fits from \textit{Brunton, 2013}). (C) Each rat's noise level and discounting rate compared to the optimal trade-off. (D) Each rat's evidence discounting parameter compared to the accuracy maximizing discounting level. (E) The average accuracy for the model fit to each rat's behavior, and optimized to maximize accuracy.}
\end{figure}

The model was fit to individual rats by maximizing the likelihood of observing the rat's choice on each trial. To evaluate the model, we can compare the reverse correlation curves from the model and subject. Figure 5A shows the comparison for an example rat, showing that the model captures the timescale of evidence discounting seen by the reverse correlation analysis. See the supplemental materials for residual error plots for each rat. 

In order to analyze the model fits we can examine the best fit parameters for each rat, and compare them to rats trained on the static version of the task (from \citealt{brunton2013}). The evidence discounting strength parameter $\lambda$ shows a striking difference between the two rat populations (Figure 5B). In the static task, the rats have small discounting rates indicating an integration timescale comparable to the longest trial the rats experienced \citep{brunton2013,hanks2015,erlich2014}. In the dynamic task, the rats have strong evidence discounting, consistent with the reverse correlation analysis. See the supplemental materials for  a comparison of other model parameters.

To assess whether rats individually calibrate their discounting timescales to their level of sensory noise, we estimate the sensory noise level from the model parameters. We estimated the click mislocalization probability by taking the average level of adaptation, and the Gaussian distributed sensory noise. Figure 5C shows each rat's fit compared to the numerically obtained optimal discounting levels from Figure 2F. The rats appear to have slightly larger discounting rates than predicted by the normative theory. The deviation from the normative theory may be due to other parameters in the behavioral model, the fact that we considered only the average level of sensory adaptation, or other factors. In order to more directly examine whether the rats were adopting the optimal timescale, we asked whether the rat's discounting rates were constrained by the other model parameters. For each rat, we took the best fitting model parameters, and froze all parameters except the discounting rate parameter $\lambda$. Then, we found the value of $\lambda$ that maximized accuracy on the trials each rat performed. Note this optimization did not ask to maximize the similarity to the rat's behavior. We found that given the other model parameters, the accuracy maximizing discounting level was very close to the rat's discounting level (Figure 5D) meaning that different sources of noise parametrized in the model highly constrain the rats' discounting rates. Further, while the discounting rates changed slightly, the improvement in total trial accuracy changed even less. For all rats, optimizing the discounting rate increased the total accuracy of the model by less than 1$\%$ (Figure 5E). Taken together these results suggest that rats discount evidence at the optimal level given several sources of noise. \newline

\noindent{\textbf{Individual rats in different environments}} \newline
Previous studies have demonstrated that rats can optimally integrate evidence in a static environment (Brunton 2013). Here we have demonstrated that rats can optimally integrate and discount evidence in a dynamic environment. In order to demonstrate the ability of individual rats to adapt their timescales in different environments, we moved three rats from a dynamic environment ($h = 0.5$ Hz) to a static environment ($h = 0$ Hz), and then back. The rats trained in each environment for many daily sessions (minimum 25 sessions). In each environment, we quantified their behavior using reverse correlation methods. Figure 6A-C show the reverse correlation curves for an example rat as the rat transitioned between environments with different statistics. Figure 6D shows the integration timescales for each rat in each environment. Rats rapidly adjusted their timescales when moving into a static environment, a session-by-session estimate is in the supplementary materials Figure 23. Consistent with our normative theory, rats in the $h = 0.5$ Hz environment show discounting rates approximately half the strength of rats in the $h = 1$ Hz environment.  We find rats can dynamically adjust their integration behavior to match their environments. \newline
\begin{figure}
\centering
\includegraphics[width=\textwidth]{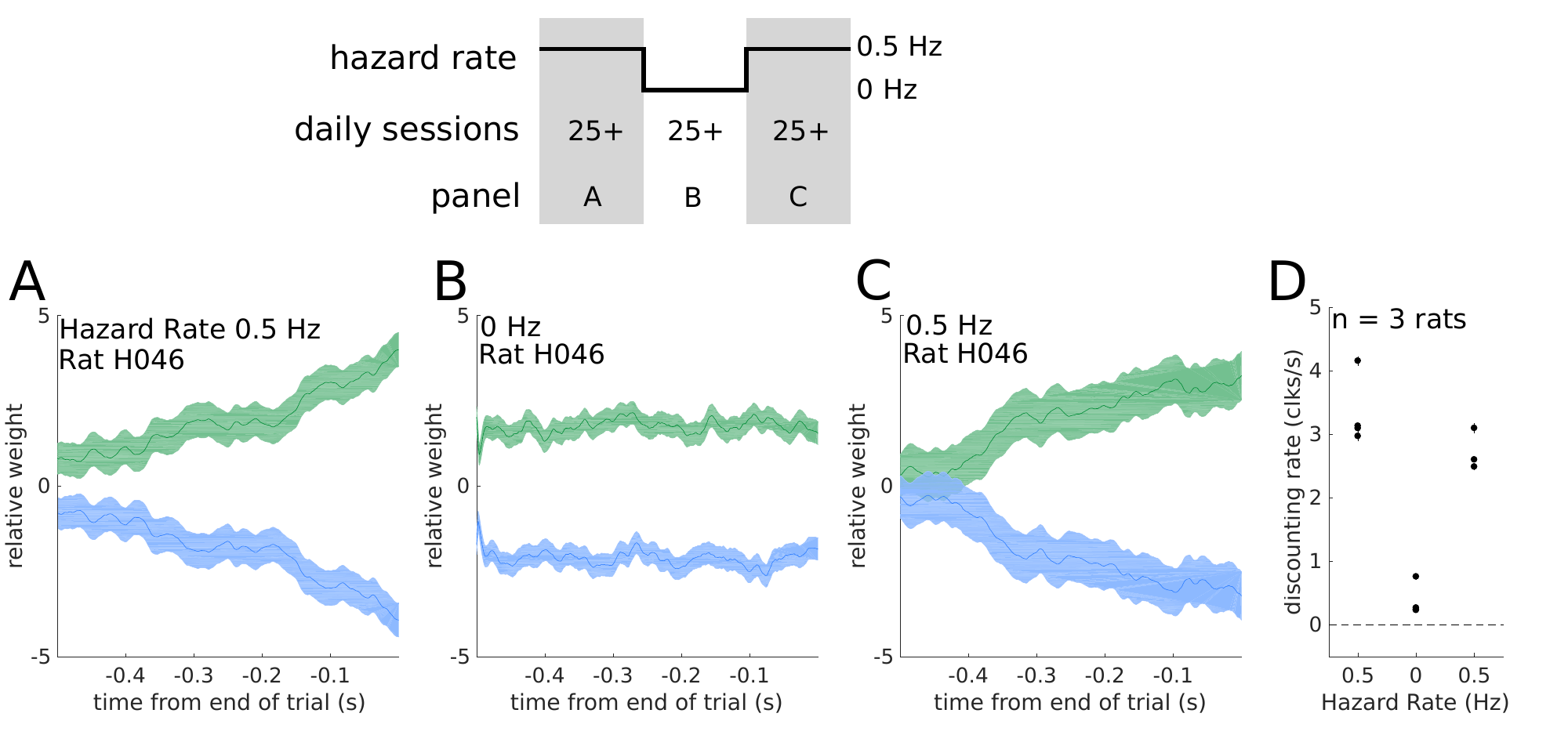}
\caption{\textbf{Rats adapt to changing environmental conditions.} Three rats were moved from a 0.5 hz hazard rate to 0 hz, then back to 0.5 hz. Rats stayed in each environment for multiple daily training sessions, with a minimum of 25 sessions. (Top) Schematic outlining the experimental design. (A-C) Reverse correlation curves for an example rat in a (A) 0.5 Hz hazard rate environment before switching, (B) 0 Hz environment, and (C) 0.5 Hz environment after switching. (D) Quantification of the integration timescales before, during, and after the switch for all rats.}
\end{figure}

\noindent {\Large \textbf{Discussion}}\newline
We have developed a pulse-based auditory decision making task in a dynamic environment. Using a high-throughput automated rat training, we trained rats to accumulate and discount evidence in a dynamic environment. Extending results from the literature \citep{kilpatrick2015}, we formalized the optimal behavior on our task, which critically involves discounting evidence on a timescale proportional to the environmental volatility and the reliability of each click. The reliability of each click depends on the experimenter imposed click statistics, and each rat's sensory noise. We find that once sensory noise is taken into account, the rats have timescales consistent with the optimal inference process. We used quantitative modeling to investigate rat to rat variability, and to predict a moment-by-moment estimate of the rats' accumulated evidence. Finally, we demonstrated rats can rapidly adjust their discounting behavior and respectively their integration timescales in response to changing environmental statistics. Our findings open new questions into complex rodent behavior and the underlying neural mechanisms of decision making. \newline

Previously accumulation of evidence has been studied in a static stationary environment . These studies have given behavioral and neural insights into the ability of rats, monkeys, and humans to optimally accumulate evidence over extended timescales \citep{brunton2013,kira2015,purcell2010neurally,philiastides2011causal,lee2004evidence,kelly2013internal,gold2001neural}.  
These studies have showed that rats or primates, like humans, can gradually accumulate evidence for decision-making, and that their evidence accumulation process timescale is optimal. Quantitative modeling revealed that errors originated from sensory noise, not from the evidence accumulation process. The optimal strategy in the stationary environment is perfect integration. A natural extension of the static version of the task is a setting in which the environment changes with some defined statistics and this what we aimed to do in our "dynamic clicks task". In the dynamics clicks task, the optimal strategy involves discounting evidence at a rate proportional to the volatility of the environment and the reliability of each evidence pulse. The behavioral quantitative modeling builds on a study that derived ideal observer models for dynamic environments, including the two-state environments considered here, and more complex environments \citep{kilpatrick2015}. That study analyzed the behavior of ideal agents with Gaussian distributed evidence samples. Our work builds on their derivation of ideal behavior, and extends their analysis to discrete evidence. Importantly, our analysis allowed us to separate evidence reliability into experimenter imposed stimulus statistics and sensory noise. Moreover, our findings show that rats discounting rates are optimal only when factoring in sensory noise. We have also shown that rats can switch back and forth between environments with  different volatilities thus providing for the first time a knob for the experimenter to control the subjects' integration timescale. 

On the other hand, a recent study examined human decision making in a dynamic environment \citep{glaze2015}. That study found that humans show nonlinear evidence discounting, but their discounting rates did not match with the optimal inference. Incorporating models of human sensory noise could explain deviations from optimality in their data. We did not examine whether rats demonstrate nonlinear evidence discounting because the linear approximation in our task is effectively indistinguishable from the full nonlinear theory (Figure 2). Other studies in humans have also found that humans perform leaky integration in dynamic environments \citep{ossmy2013}.

The behavior presented here is distinct from previous tasks that have investigated decision making over time. \citealt{Cisek11560} developed an evidence accumulation task in which the amount of evidence changes over the course of the trial. However, in that study the evidence is generated from a stationary process and the optimal behavior is to perfectly integrate all evidence. This is in contrast to the present study that examines conditions under which the optimal behavior is to discount old evidence. 

In a separate line of work called bandit tasks, the subject gets reward or feedback on a timescale slower than the dynamics of the environment \citep{Iigaya141309, miller2017}. In bandit tasks, the environment changes slowly with respect to each choice, and subjects get many opportunities for reward and feedback before the environment changes. In the work presented here, the subjects must perform inference without feedback while the dynamics of the environment are changing within the course of one trial. Importantly, in our task the environmental state ``resets'' after each choice the rat makes. 

The dynamic accumulation of evidence task that we are presenting here should not also be confused with the conventional change detection tasks, which have only a single change of mind. In our case, we have many changes of mind that are happening stochastically. See Fig 2 in \citealt{kilpatrick2015} for a detailed discussion on the relationship between these tasks. 

It is very important to note that the term ``evidence discounting'' is different than ``temporal discounting'' prominently used in the reinforcement learning literature. Temporal discounting is the phenomenon in which the subjective value of some reward decreases in magnitude when the given reward is delayed \citep[pg.352]{dayanAbbott}.  In our case, evidence discounting is the phenomenon in which an agent discards evidence in order to infer state changes in the environment.

One benefit of rodent studies is the wide range of experimental tools available to investigate the neural mechanisms underlying behavior. Our task will facilitate the investigation of two neural mechanisms. First, due to the dynamic nature of each trial, subject's change their mind often during each trial allowing experimental measurement of changes of mind within one trial. Further, these changes of mind are driven by internal estimates of accumulated evidence. Previous studies of rat decision making have identified a cortical structure, the Frontal Orienting Fields (FOF) as a potential substrate for upcoming choice memory \citep{erlich2011, hanks2015,erlich2014,kopec2014,piet2017}. Future work could investigate if and how the FOF tracks upcoming choice in a dynamic environment during changes of mind. It will also complement already existing neurophysiological studies of changes of mind \citep{kiani2014, peixoto2016}

Second, normative behavior in a dynamic environment requires tuning the timescale of evidence integration to the environmental volatility. There is a large body of experimental and theoretical studies on neural integrators \citep{seung1996,goldman2008,Aksay2007, scott2017} that investigates how neural circuits potentially perform integration. Many possible neural circuit mechanisms have been proposed, from random unstructured networks \citep{maass2002,ganguli2008}, feed-forward syn-fire chains \citep{goldman2008}, and recurrent structured networks of many forms \citep{seung1996, druckmann2012, boerlin2013}. The task developed here allows for experimental control of the putative neural integrator's timescale within the same subject. Measurement of neural activity in different dynamic environments, and thus different integration timescales, may shed light into which mechanisms are used in neural circuits for evidence integration. For instance, unstructured networks, or feed-forward networks may re-tune themselves via adjusting read-out weights. Networks that integrate via recurrent dynamics; however, would re-tune themselves via changes in those recurrent dynamics. Alternatively, measurement of neural activity in different dynamic environments may reveal fundamentally new mechanisms of evidence integration. For instance, \citealt{erlich2014} proposed multiple integration networks with different timescales to account for behavioral changes in response to prefrontal cortex inactivations. Our task may allow further investigation into the structure and dynamics of neural integrators.  \newline

\noindent {\Large \textbf{Methods}}\newline
\noindent{\textbf{Subjects}}\newline
Animal use procedures were approved by the Princeton University Institutional Animal Care and Use Committee and carried out in accordance with NIH standards. All subjects were adult male Long Evans rats (Vendor: Taconic and Harlan, USA) placed on a controlled water schedule to motivate them to work for a water reward. 
\newline 

\noindent{\textbf{Behavioral Training}}\newline
We trained 14 rats on the dynamic clicks task (Figure 1). Rats went through several stages of an automated training protocol. In the final stage, each trial began with an LED turning on in the center nose port indicating to the rats to poke there to initiate a trial. Rats were required to keep their nose in the center port (nose fixation) until the light turned off as a ``go'' signal. During center fixation, auditory cues were played indicating the current hidden state. The duration of the fixation period (and stimulus period) ranged from 0.5 to 2 seconds. After the go signal, rats  were rewarded for entering the side port corresponding to the hidden state at the end of the stimulus period. The hidden state did not change after the go signal. A correct choice was rewarded with 24 microliters of water; while an incorrect choice resulted in a punishment noise (spectral noise of 1 kHz for a 0.7 seconds duration). The rats were put on a controlled water schedule where they receive at least 3\% of their weight every day.  Rats trained each day in a training session on average 120 minutes in duration. Training sessions were included for analysis if the overall accuracy rate exceeded 70\%, the center-fixation violation rate was below 25\%, and the rat performed more than 50 trials. In order to prevent the rats from developing biases towards particular side ports an anti-biasing algorithm detected biases and probabilistically generated trials with the correct answer on the non-favored side. \newline 

\noindent{\textbf{Linear discounting agents }}\newline
To analyze the performance of linear discounting agents at varying levels of noise, we created synthetic noisy-datasets. For each level of click noise, each click switched sides according to the noise level. On each of these datasets, we numerically optimized the discounting level that maximized the accuracy of predicting the hidden state at the end of the trial. \newline

\noindent{\textbf{Psychophysical reverse correlation}}\newline
The computation of the reverse correlation curves was very similar to methods previously reported \citep{brunton2013,hanks2015,erlich2014}. However, one additional step is included to deal with the hidden state. The first step is to smooth the click trains on each trial with a causal Gaussian filter ($k(t)$), this creates one smooth click rate for each trial. The filter had a standard deviation of 5 msec. 
\begin{eqnarray}
r_i(t) = \delta_{R,t}\ast k(t) - \delta_{L,t} \ast k(t) 
\end{eqnarray} Then, the smooth click rate on each trial was normalized by the expected click rate for that time step, given the current state of the environment. This gives us the deviation (the excess click rate) from the expected click rate for each trial.
\begin{eqnarray}
e_i(t) = r_i(t) - \langle r(t) | S_i(t) \rangle
\end{eqnarray} Finally, we compute the choice triggered average of the excess click rate by averaging over trials based on the rat's choice. 
\begin{eqnarray}
\text{excess-rate}(t|\text{choice}) = \langle e(t) | \text{choice} \rangle
\end{eqnarray}
The excess rate curves were then normalized to integrate to one. This was done to remove distorting effects of a lapse rate, as well to make the curves more interpretable by putting the units into effective weight of each click on choice. To quantify the timescale of the reverse correlation curves, we fit an exponential of the form $ae^{bt}$ to each curve. The parameter $b$ is the discounting rate, while $1/b$ is the integration timescale. \newline

\noindent{\textbf{Behavioral Model}}\newline
Previous studies using this behavioral accumulation of evidence model \citep{brunton2013} have included sticky bounds which absorb probability mass when the accumulated evidence reaches a certain threshold. We found this sticky bounds to be detrimental to high performance on our task, so we removed them. The removal of the sticky bounds facilitates an analytical solution of the model. The model assumes an initial distribution of accumulation values $P(a | t=0) = \mathcal{N} \left(\mu_0, \sigma^2_i\right)$. At each moment in the trial, the distribution of accumulation values $P(a|t,\delta_R, \delta_L)$ is Gaussian distributed with mean ($\mu$) and variance  ($\sigma^2$) given by:
\begin{eqnarray}
\mu(t) &=& \mu_0 e^{\lambda t} + \int\limits_0^t \left(\delta_{R,s}\cdot C\left(R(s)\right) - \delta_{L,s}\cdot C\left(L(s)\right) \right)ds \\
\mu(t) &=& \mu_0 e^{\lambda t} + \sum\limits_i^{\#R} e^{\lambda\left(t-R(i)\right)}C(R(i)) - \sum\limits_i^{\#L} e^{\lambda\left(t-L(i)\right)}C(L(i))\\
\sigma^2(t) &=& \sigma_i^2e^{\lambda t} + \frac{\sigma_a^2}{2\lambda}\left( e^{2 \lambda t} - 1 \right) + \int\limits_0^t \sigma_s^2 \left(\delta_{R,s}\cdot C\left(R(s)\right) - \delta_{L,s}\cdot C\left(L(s)\right)  \right) e^{2\lambda t} ds \\
\sigma^2(t) &=& \sigma_i^2e^{\lambda t} + \frac{\sigma_a^2}{2\lambda}\left( e^{2 \lambda t} - 1 \right) + \sum\limits_i^{\#R} \sigma_s^2C(R(i))e^{2\lambda \left(t - R(i)\right)} + \sum\limits_i^{\#L} \sigma_s^2C(L(i))e^{2\lambda \left(t - L(i)\right)}
\end{eqnarray} Where $\#R$ is the number of right clicks on this trial up to time $t$, and $R(i)$ is the time of the $i^{th}$ right click. $C(R(i))$ tells us the effective adaptation for that clicks. For a detailed discussion of a similar model, see \citealt{feng2009}.

Given a distribution of accumulation values $P(a|t,\delta_R, \delta_L) = \mathcal{N}\left(\mu(t), \sigma^2(t)\right)$, and the bias parameter $B$, we can compute the left and right choice probabilities by:
\begin{eqnarray}
P(\text{go right}) &=& \frac{1}{2}\left(1+ \text{erf}\left( \frac{-\left(B-\mu(t) \right)}{\sigma \sqrt{2}} \right) \right), \\
P(\text{go left}) &=& 1 - P(\text{go right}).
\end{eqnarray}
These choice probabilities are then distorted by the lapse rate, which parameterizes how often a rat makes a random choice:. The model parameters $\theta$ were fit to each rat individually by maximizing the likelihood function:
\begin{eqnarray}
L = \prod_i^{\#\text{trials}}P(\text{rat's choice on trial i}| \theta, \delta_R^i, \delta_L^i).
\end{eqnarray} 
Additionally, a half-gaussian prior was put on the initial noise ($\sigma_i$) and accumulation noise parameters ($\sigma_a$). Due to the presence of large discounting rates, these parameters are difficult to recover in synthetic datasets. The priors were set to match the respective best fit values from \citealt{brunton2013}. The numerical optimization was performed in MATLAB. To estimate the uncertainty on the parameter estimates, we used the inverse hessian matrix as a parameter covariance matrix \citep{daw2011}. To compute the hessian of the model, we used automatic differentiation to exactly compute the local curvature \citep{autodiff}. \newline

\noindent{\textbf{Calculating noise level from model parameters}}\newline
Given the model parameters $(\sigma^2_s$, $\phi$, and $\tau_{\phi})$, we computed the average level of sensory adaptation on each click $\langle C \rangle$. Then, we computed what fraction of the probability mass would cross 0 to be registered as a click on the other side.
\begin{eqnarray}
n &=& \frac{1}{2}\left(1+erf\left(\frac{-\langle C \rangle}{\sqrt{2 \sigma^2_s \langle C \rangle }} \right) \right).
\end{eqnarray}
\newline

\noindent {\Large \textbf{Acknowledgements}}\newline
We thank all members of the Brody lab for technical assistance, and feedback throughout the project. We thank Ben Scott, Diksha Gupta, Tim Hanks, and Christine Constantinople for detailed comments on the manuscript. This work was supported in part by NIH grant 5-R01-MH108358 \newline

\noindent {\Large \textbf{Author Contributions}}\newline
AP:Task design, rat training, theoretical analysis, quantitative methods development and application. AE: Task design, rat training, and advised during all aspects of the study. CB: advised during all aspects of the study

\newpage
\noindent {\LARGE \textbf{Supplementary Materials}}\newline
The supplementary materials contains extended figures and control analyses for several aspects of the study: 
\begin{enumerate}
\item Individual rat behavior and training
\item Optimal inference details and derivation
\item Sensory noise parameterization details and alternatives
\item Psychophysical reverse correlation details
\item Quantitative model details \newline
\end{enumerate}

\noindent{\Large\textbf{Individual Rat Behavior}}\newline
In this section we outline our rat training process, and then provide several model free analyses of behavior for each rat individually. First, we include the psychometric curve with respect to the total click difference on each trial. Second, we include the psychometric curve with respect to the optimal inference process (assuming no sensory noise). Third, the chronometric plot shows rat accuracy with respect to time since the last hidden state change. Fourth, the chronometric plot with respect to total trial duration. Fifth, the reverse correlation curves with best fit exponential for each rat. \newpage

\noindent{\Large\textbf{Training procedure details}}\newline
The training process for the Dynamic Clicks task involves ``classical'' training during which the rats learn to associate ports with rewards. The ``classical'' pipeline is the same as \citealt{Duan20151491}. After completing classical training, then rats then learn the standard Poisson clicks task as described in \citealt{brunton2013}. Finally, environmental state switching is introduced. The following tables outline keep changes at each stage in the procedure. Rats typically spend a few days on classical, and about 6 weeks moving through the clicks training. \newline 
\textbf{Classical Training:}

\begin{tabular}{|c | l | c | }
\hline
 &Stage Name                                \\ \hline
1 &learn left poking                        \\
2 &learn right poking                       \\
3 &learn center poke switching blocks       \\
\hline
\end{tabular}\newline

\textbf{Dynamic Clicks Training:}

\begin{tabular}{|c | l | c | c | c |c | }
\hline
 &Stage Name                       &h  	& $\gamma$  & T & $\Delta T$     \\ \hline
1 &grow nose in center 1           &0  	& 5   		&1  & 0       \\
2 &wait for good endpoints         &0  	& 5   		&1  & 0       \\
3 &grow nose in center 2           &0  	& 5   		&2  & 0       \\
4 &wait for good endpoints         &0  	& 5   		&2  & 0        \\
5 &add variable trial length       &0  	& 5   		&2  & 0.5      \\
6 &wait for good endpoints         &0  	& 5   		&2  & 0.5      \\
7 &step hazard 1                   &1 Hz& 5   		&2  & 0.5      \\
9 &increase variable trial length  & 1  & 5  		&2  & 1.5      \\
10 &add psychometrics 1            & 1  & 4  		&2  & 1.5      \\
11 &add psychometrics 2            & 1  & 3.5		&2  & 1.5      \\
12 &add psychometrics 3            & 1  & 3.2		&2  &1.5       \\
13 &add psychometrics 4            & 1  & 3  		&2  & 1.5      \\
14 &final                          & 1  & 3  		&2  & 1.5      \\
\hline
\end{tabular}\newline
\noindent\textbf{Notes: }
\textbf{h} is the hazard rate between states. H is introduced in stage 9, increased in stage 10, and is constant afterwards. \textbf{Gamma ($\gamma$)} is the log of the click rates. $\gamma = 5$ is the setting for endpoints, which
 is extremely easy. \textbf{T} is the maximum duration of the stimulus period. \textbf{$\Delta T$} is the variability in stimulus period duration. If the trial length (T) is  2 seconds, and $\Delta T$ is 0.5, that means trials are 1.5 - 2 seconds in length. \newpage

\clearpage
\begin{figure}
\includegraphics[width=\textwidth]{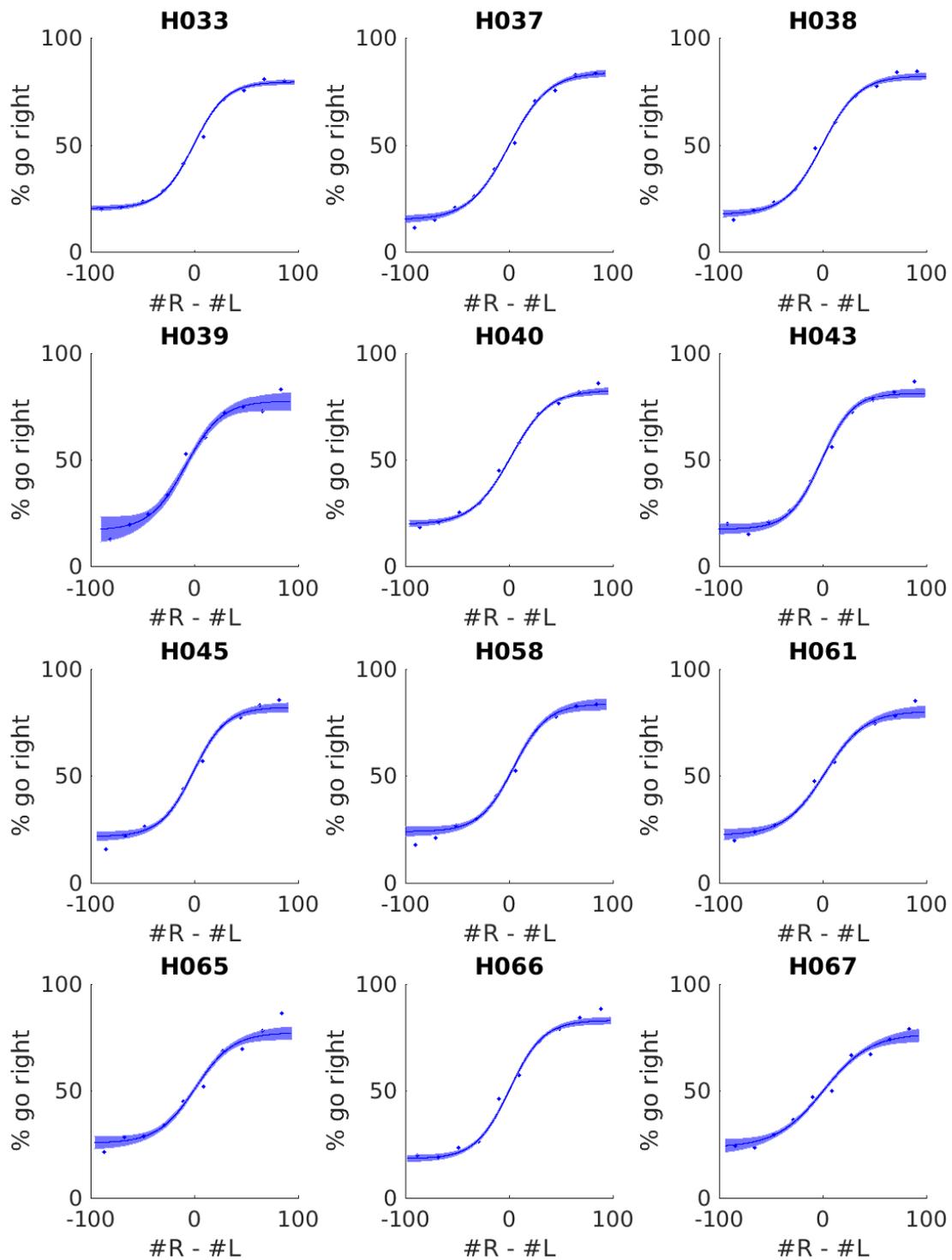}
\caption{\textbf{Psychometric graph for all rats} Each trial performed by the rat was binned by the total click difference in the trial. The rat's average accuracy in each bin is shown (dots). A four parameter logistic function is fit to the data with 95\% confidence intervals (line).}
\end{figure}
\begin{figure}
\includegraphics[width=\textwidth]{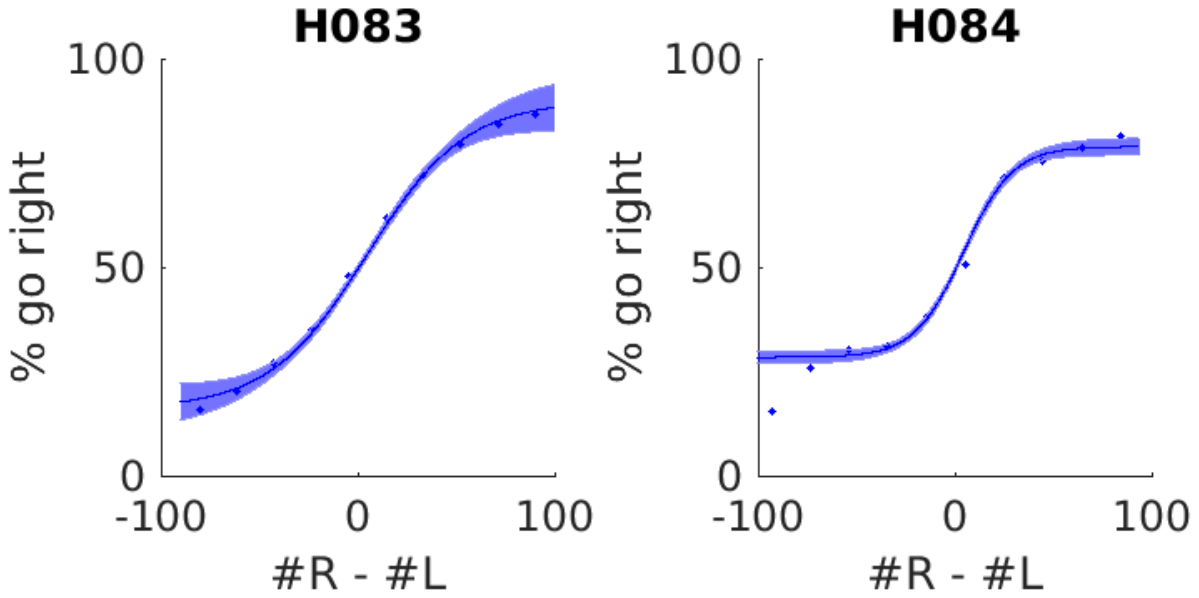}
\caption{\textbf{Psychometric graph for all rats} Each trial performed by the rat was binned by the total click difference in the trial. The rat's average accuracy in each bin is shown (dots). A four parameter logistic function is fit to the data with 95\% confidence intervals (line).}
\end{figure}
\begin{figure}
\includegraphics[width=\textwidth]{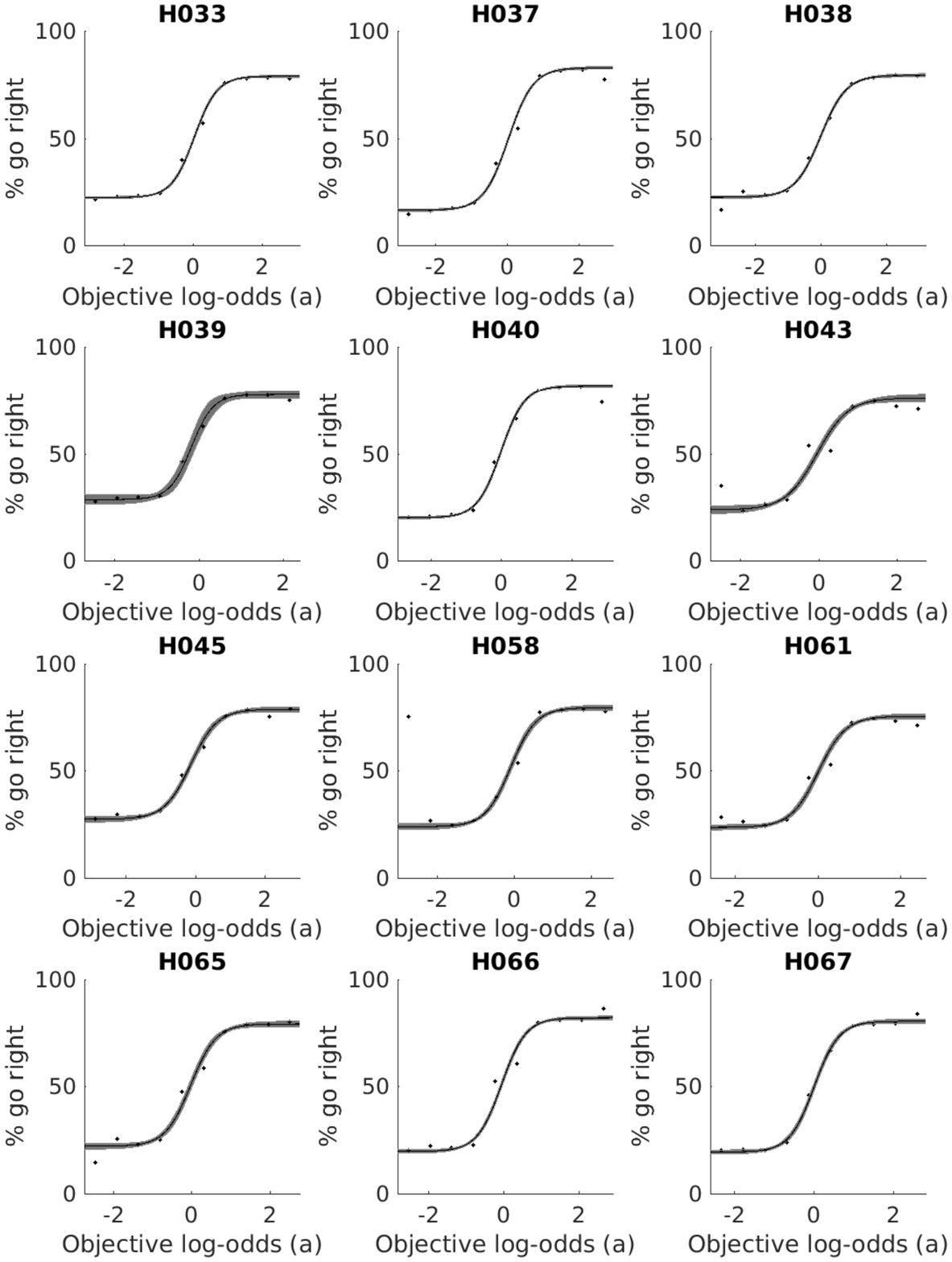}
\caption{\textbf{Psychometric graph for all rats against ideal observer} Each trial performed by the rat was binned by the accumulation value (log-odds) of the ideal observer (ie, no sensory noise). The rat's average accuracy in each bin is shown (dots). A four parameter logistic function is fit to the data with 95\% confidence intervals (line).}
\end{figure}
\begin{figure}
\includegraphics[width=\textwidth]{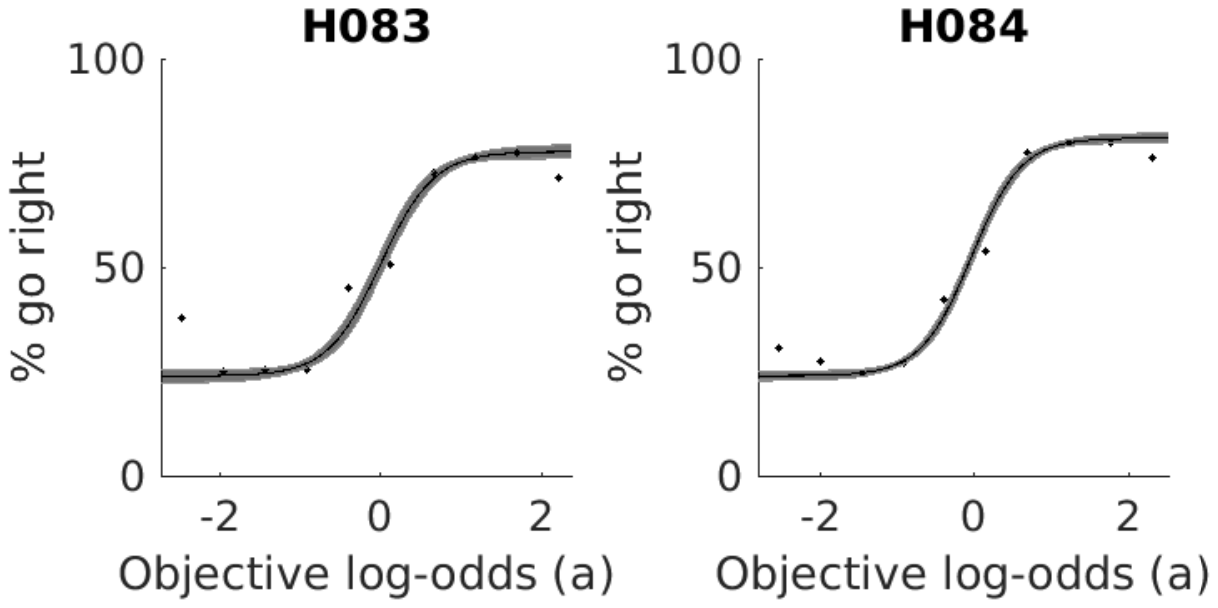}
\caption{\textbf{Psychometric graph for all rats against ideal observer} Each trial performed by the rat was binned by the accumulation value (log-odds) of the ideal observer (ie, no sensory noise). The rat's average accuracy in each bin is shown (dots). A four parameter logistic function is fit to the data with 95\% confidence intervals (line).}
\end{figure}
\begin{figure}
\includegraphics[width=\textwidth]{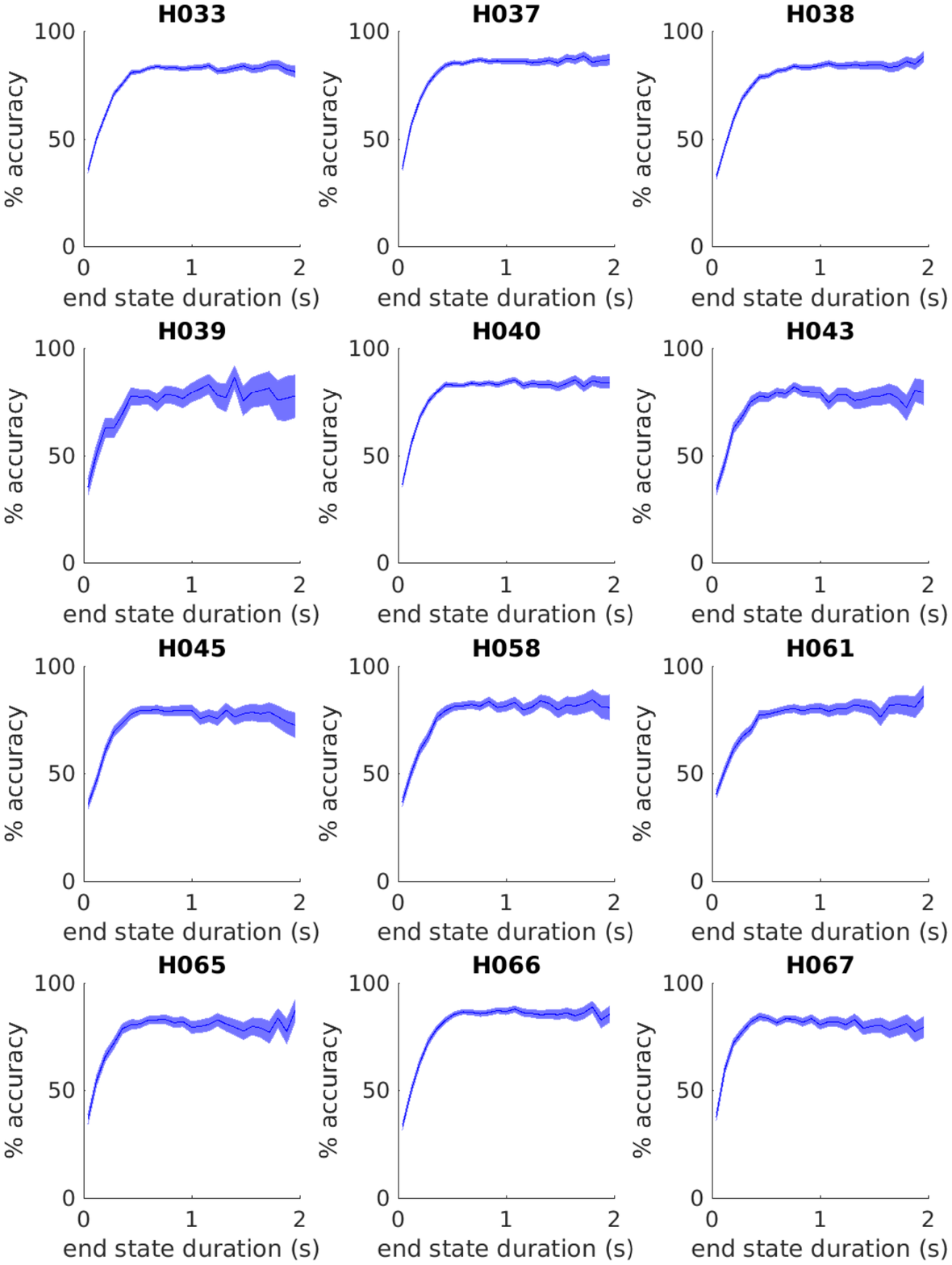}
\caption{\textbf{Chronometric graph for all rats} Each trial was binned by the amount of time since the last change in the hidden environmental state. The average accuracy of each bin is shown.}
\end{figure}
\begin{figure}
\includegraphics[width=\textwidth]{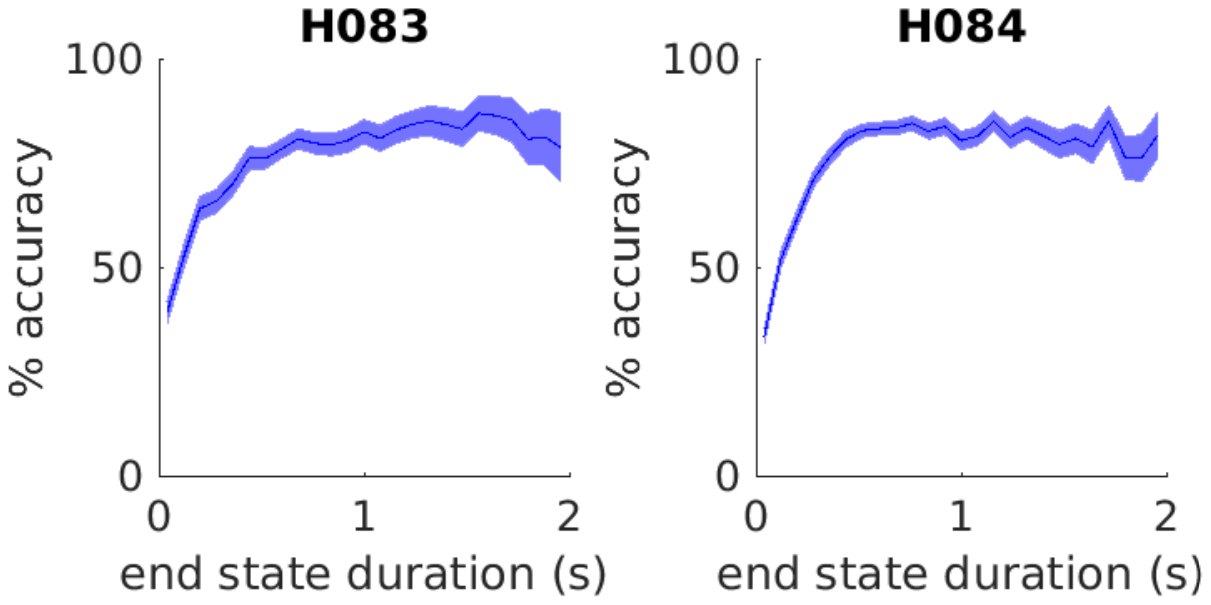}
\caption{\textbf{Chronometric graph for all rats} Each trial was binned by the amount of time since the last change in the hidden environmental state. The average accuracy of each bin is shown.}
\end{figure}
\begin{figure}
\includegraphics[width=\textwidth]{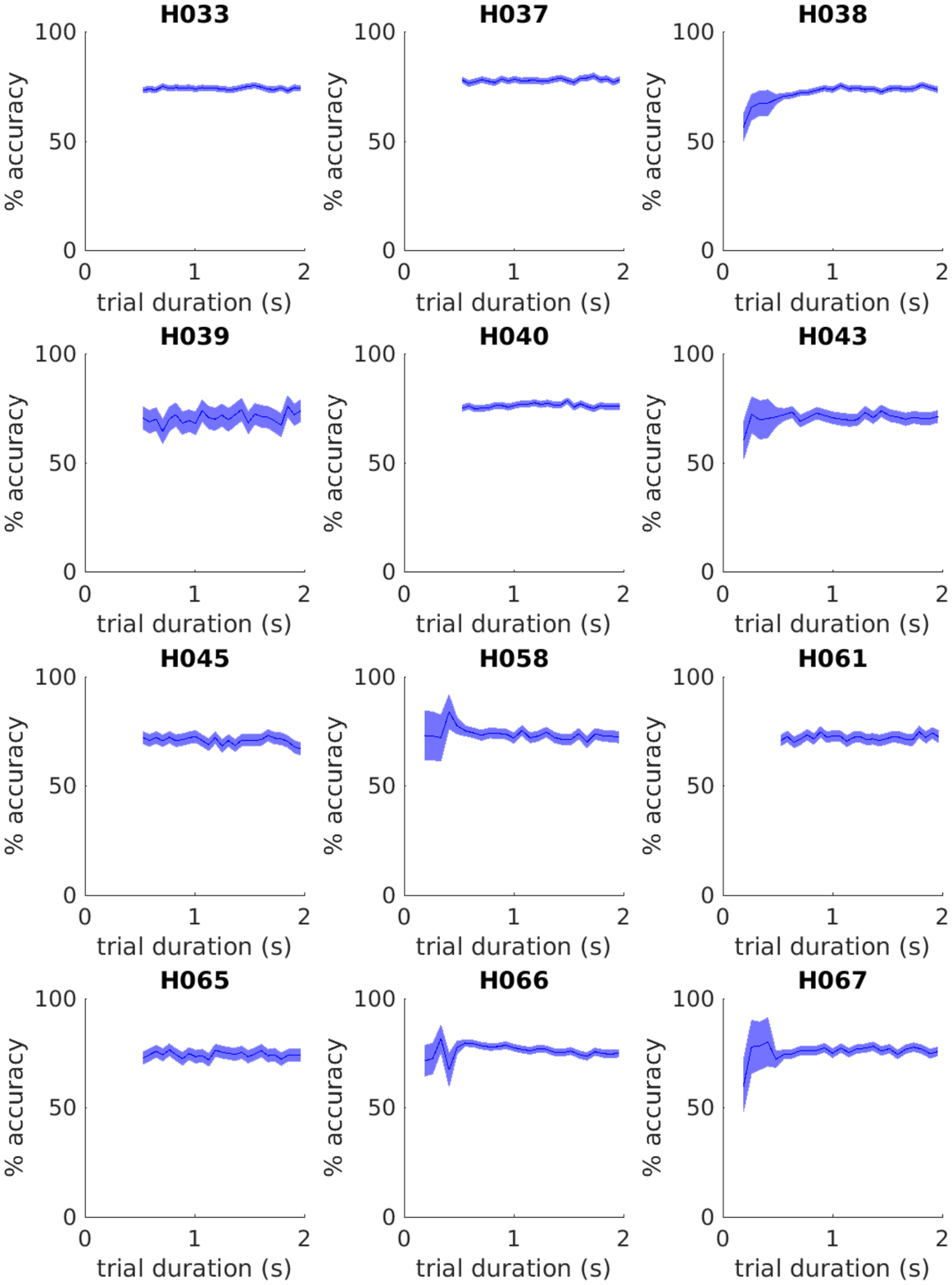}
\caption{\textbf{Chronometric graph for all rats} Each trial was binned by the total trial duration. The average accuracy of each bin is shown. Most trials were drawn from the range (0.5 - 2) seconds; however, some rats experience a small number of shorter trials, leading to greater uncertainty for those durations.}
\end{figure}
\begin{figure}
\includegraphics[width=\textwidth]{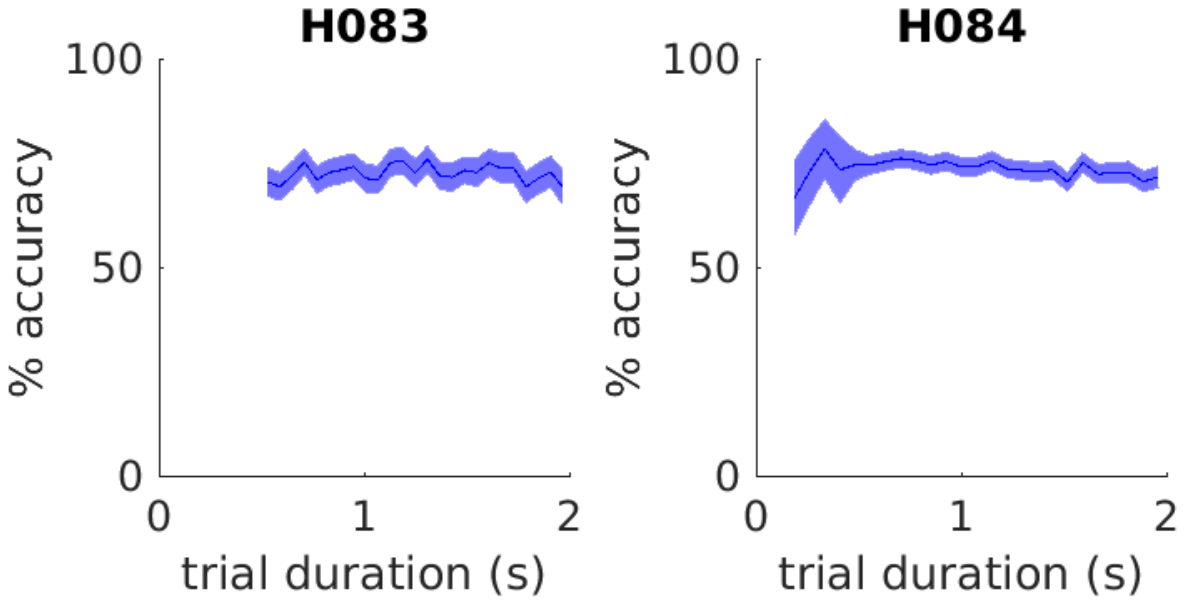}
\caption{\textbf{Chronometric graph for all rats} Each trial was binned by the total trial duration. The average accuracy of each bin is shown. Most trials were drawn from the range (0.5 - 2) seconds; however, some rats experience a small number of shorter trials, leading to greater uncertainty for those durations.}
\end{figure}
\begin{figure}
\includegraphics[width=\textwidth]{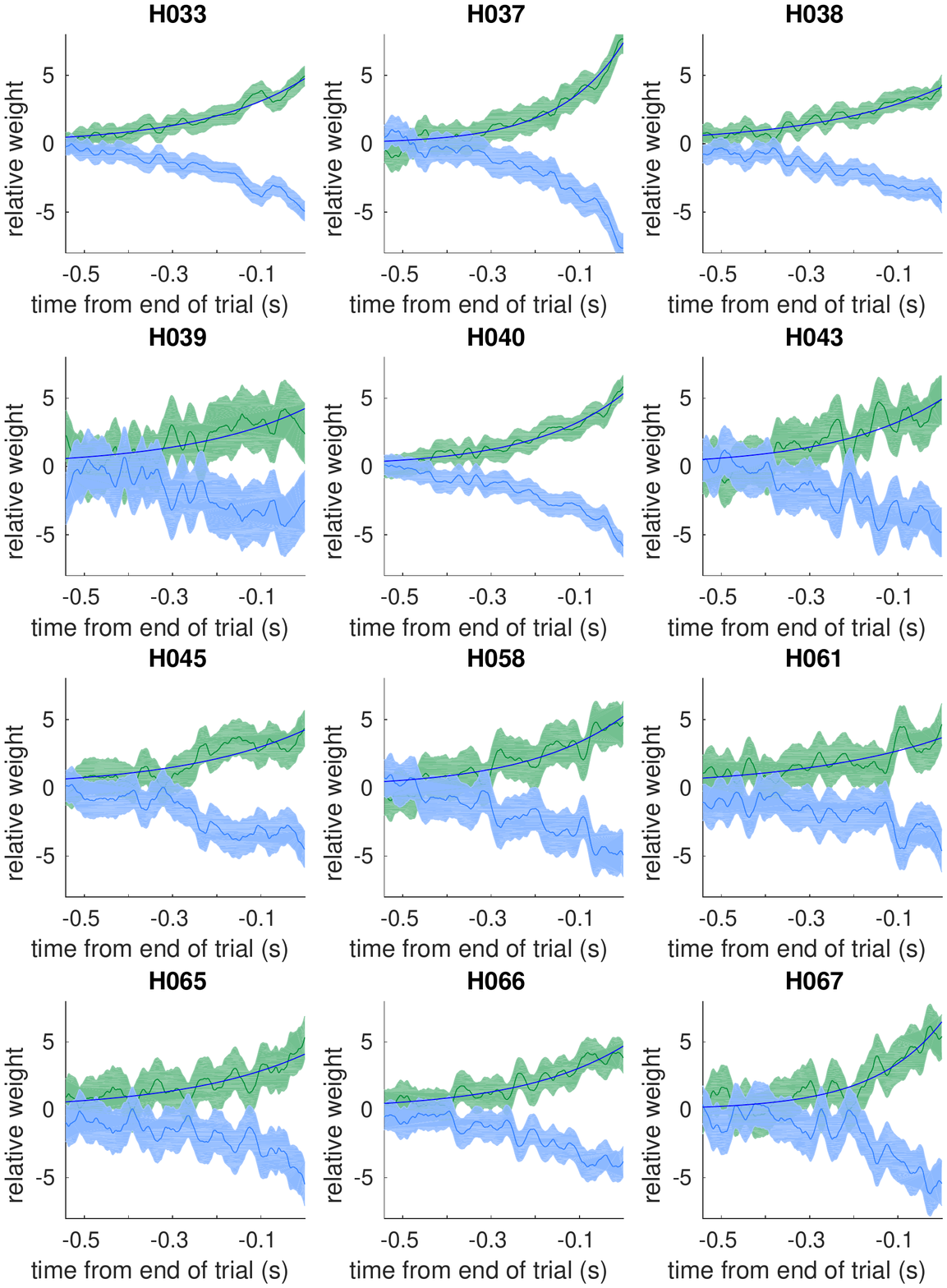}
\caption{\textbf{Reverse Correlation for all rats} Reverse Correlation curves for each rat (black), as well as the best fit exponential discounting function (blue).}
\end{figure}
\begin{figure}
\includegraphics[width=\textwidth]{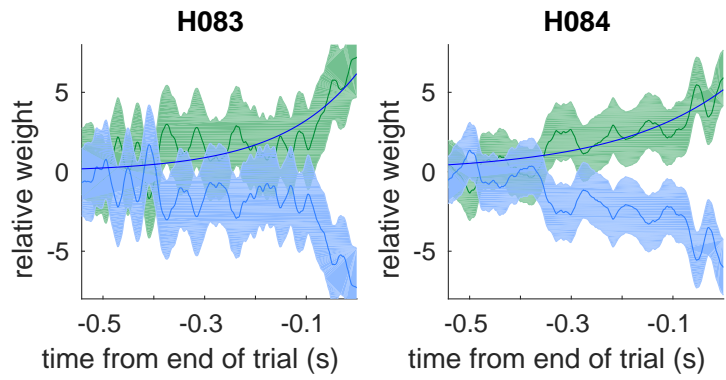}
\caption{\textbf{Reverse Correlation for all rats} Reverse Correlation curves for each rat (black), as well as the best fit exponential discounting function (blue).}
\end{figure}\newpage 

\clearpage
\noindent{\Large\textbf{Optimal inference details}}\newline
\noindent{\textbf{Derivation of optimal inference}}\newline
Here we provide more detail on the derivation from equation \eqref{ratio} to equation \eqref{log_ratio}. This derivation was developed by \citealt{kilpatrick2015}, see equations 3.2 and 3.3. However, we do not approximate the evidence term into its first two moments, instead evaluating the evidence term. For this reason we report the same derivation but halting at the intermediate step not shown in \citealt{kilpatrick2015}. \newline

\noindent Beginning with the evidence ratio, equation \eqref{ratio} in the present study, and equation 3.2 in \citealt{kilpatrick2015}.
\begin{align}
R_t &=\frac{P(S^1 | \epsilon_{1 \ldots t})}{P(S^2| \epsilon_{1 \ldots t})} = \frac{P(\epsilon_t | S^1)}{P(\epsilon_t | S^2)}\left(\frac{\left(1-h\Delta t \right)R_{t-1} + h\Delta t  }{\left(h\Delta t\right) R_{t-1} + 1 -h \Delta t  }\right). \\
\intertext{Dividing each side by $R_{t-1}$}
\frac{R_t}{R_{t-1}} &= \frac{P(\epsilon_t | S^1)}{P(\epsilon_t | S^2)}\left(\frac{\left(1-h\Delta t \right)R_{t-1} + h\Delta t  }{\left(h\Delta t\right) R_{t-1} + 1 -h \Delta t  }\right)\frac{1}{R_{t-i}}. \\ 
\intertext{Now, define $\hat{a}_t = \log\left(R_t\right)$, and take the logarithm of both sides:}
\hat{a}_t - \hat{a}_{t-1} &= \log\left(\frac{P(\epsilon_t | S^1)}{P(\epsilon_t | S^2)}\right) + \log\left(\frac{\left(1-h\Delta t \right) + h\Delta te^{-\hat{a}_{t-1}}  }{\left(h\Delta t\right) e^{\hat{a}_{t-1}} + 1 -h \Delta t  }\right), \\
\Delta \hat{a}_t &= \log\left(\frac{P(\epsilon_t | S^1)}{P(\epsilon_t | S^2)}\right) + \log\left(\frac{\left(1-h\Delta t \right) + h\Delta te^{-\hat{a}_{t-1}}  }{ \left( 1 - h\Delta t\right) + h\Delta t e^{\hat{a}_{t-1}} }\right), \\
\Delta \hat{a}_t &= \log\left(\frac{P(\epsilon_t | S^1)}{P(\epsilon_t | S^2)}\right) + \log\left( 1 + h\Delta t  \left(te^{-\hat{a}_{t-1}} - 1 \right)\right) - \log\left(1 + h\Delta t \left( e^{\hat{a}_{t-1}} -1\right) \right) \\ 
\intertext{Using the approximation $\log\left(1+a\right) \approx a$, which is valid when $|a| << 1$. Here, $h\Delta t << 1$.} 
\Delta \hat{a}_t &= \log\left(\frac{P(\epsilon_t | S^1)}{P(\epsilon_t | S^2)}\right) +  h\Delta t  \left(e^{-\hat{a}_{t-1}} - 1 \right) - h\Delta t \left( e^{\hat{a}_{t-1}} -1\right), \\
\Delta \hat{a}_t &= \log\left(\frac{P(\epsilon_t | S^1)}{P(\epsilon_t | S^2)}\right) +  h\Delta t  \left(e^{-\hat{a}_{t-1}} -  e^{\hat{a}_{t-1}} \right).\\ 
\intertext{Using $\sinh(x) = \frac{1}{2}\left(e^x - e^{-x}\right)$:}
\Delta \hat{a}_t &= \log\left(\frac{P(\epsilon_t | S^1)}{P(\epsilon_t | S^2)}\right) -2h\Delta t \sinh \left(\hat{a}_{t-1} \right).\\ 
\intertext{Here, we again use $\Delta t << 1$ to justify replacing $\hat{a}_{t-1}$ with $\hat{a}_{t}$ on the right hand side. Evaluating the evidence term as derived in the main text, and rescaling $\kappa$:}
\Delta a_t &= \delta_{t,R} - \delta_{t,L}  -\frac{2h}{\kappa}\Delta t \sinh \left(\kappa a_t\right).\\ 
\intertext{Taking the limit of $\Delta t \rightarrow 0 $:}
da_t &= \delta_{t,R} - \delta_{t,L}  -\frac{2h}{\kappa} \sinh \left(\kappa a_t\right) dt.
\end{align}
Here we are making the assumption that the action of the auditory clicks happen instantaneously with respect to the accumulation equation. \newline

\noindent {\textbf{Decreasing click reliability lengthens integration timescales}}\newline
We found that plotting the evidence discounting term with less reliable clicks (smaller $\kappa$) resulted in a flatter curve, which corresponds to a longer integration timescale. To see this relationship more clearly we can expand the discounting function in a taylor series around the origin:
\begin{eqnarray}
f(x) &\approx& f(a) + \frac{f'(a)}{1!}\left(x-a\right) + \frac{f''(a)}{2!}\left(x-a\right)^2 + \frac{f'''(a)}{3!}\left(x-a\right)^3 + \ldots \\
\frac{2h}{\kappa}\sinh\left(\kappa x\right) &\approx& \frac{2h}{\kappa}\sinh\left(0\right) + \frac{2h}{\kappa}\frac{\kappa}{1!}\cosh\left(\kappa\cdot  0\right)\left(x-0\right)+ \frac{2h}{\kappa}\frac{\kappa^2}{2!}\sinh\left(\kappa\cdot  0\right)\left(x-0\right)^2 
\end{eqnarray}The even terms drop out, and we collect the odd terms:
\begin{eqnarray}
\frac{2h}{\kappa}\sinh\left(\kappa x\right) &\approx& 2hx + \frac{2h\kappa^2}{3!}x^3 + \frac{2h\kappa^4}{5!}x^5 + \ldots 
\end{eqnarray} We find that $\kappa$ only appears with even power exponents in odd powers of x. Increasing $\kappa$ will increase the strength of the discounting function. Increasing the strength of the discounting function leads to shorter integration timescales. In short, increasing $\kappa$ shortens the integration timescale. Decreasing $\kappa$ lengthens the integration timescale. \newpage

\noindent{\Large \textbf{Sensory noise parameterization details }}\newline
The main analysis in the text derives optimal inference given sensory noise that is discrete, clicks are either localized on one side or the other. It is easy to imagine many other forms of sensory noise, including Gaussian fluctuations in the click amplitude, or simply missing clicks. Here we demonstrate by evaluating the log-evidence term that decreases in click reliability are primarily driven by click mislocalization, not fluctuations in the perceived amplitude of the clicks, or missed clicks. Finally, we provide a general argument for why only click mislocalization matters \newline

\noindent{\textbf{Click reliability with Gaussian sensory noise}}\newline
Consider Gaussian noise where the clicks played from the right/left are perceived with amplitude given by $\mathcal{N}(\pm\mu,\sigma^2)$. Here we interpret clicks with positive amplitude as right clicks, and negative amplitude as left clicks. Note that if $\sigma^2$ is sufficiently large, clicks will be mislocalized. 

To compute the reliability of an individual click with a specific amplitude fluctuation $(a)$, we need to compute the probability of a click generated on the right being observed with amplitude $a$: $P_r(a)$, as well as the probability of a click generated on the left being observed with amplitude $a$: $P_l(a)$. Formally we need to integrate the Gaussian probability density function over a small window centered at $a$. 

\begin{align}
P_r(a) &= \int\limits_{a-\epsilon}^{a+\epsilon} \frac{1}{\sqrt{2\pi \sigma^2}} e^{-\frac{(\mu-s)^2}{2\sigma^2}} ds \\
P_l(a) &= \int\limits_{a-\epsilon}^{a+\epsilon} \frac{1}{\sqrt{2\pi \sigma^2}} e^{-\frac{(-\mu-s)^2}{2\sigma^2}} ds \\
\kappa\left(r_1, r_2,P_r,P_l\right) &=\log\frac{(r_1\Delta t)P_r(1- r_2\Delta t)+(1-r_1\Delta t)(r_2\Delta t)P_l  }{(r_2\Delta t)P_r(1- r_1\Delta t)+(1-r_2\Delta t)(r_1\Delta t)P_l} . \\
\intertext{This expression for $\kappa$ seems hard to interpret, but notice what happens if $P_l = 0$. }
\kappa\left(r_1, r_2,P_r,0\right) &=\log\frac{(r_1\Delta t)(1- r_2\Delta t)}{(r_2\Delta t)(1- r_1\Delta t)} =\kappa(r_1,r_2). 
\end{align} In this case, $P_r$ drops out entirely, and we get the same value of $\kappa$ as the no-noise case.  This demonstrates that click mislocalization is necessary for a decrease in click reliability.  

Next, we will compare how the Gaussian click reliability scales with the rate of mislocalization. We generated a dataset of  trials where each click had an amplitude drawn from a Gaussian distribution. We asked what was the accuracy of the nonlinear inference using the Gaussian click reliability derived above, and what is the discounting rate of the best linear discounting agent? We refer to this as ``quenched Gaussian noise,'' the meaning of quenched is explained below. We then considered a second dataset where the Gaussian amplitudes were thresholded to either be $\pm 1$ reflecting whether the amplitude was above or below 0. We refer to this as ``discrete noise.'' We compute the click mislocalization probability for corresponding to each Gaussian variance $\sigma^2$ by:
\begin{eqnarray}
\langle n(\mu,\sigma^2) \rangle &= \frac{1}{2}\left(\mu+erf\left(\frac{1}{\sqrt{2 \sigma^2_s }} \right) \right)
\end{eqnarray}

\begin{figure}
\centering
\includegraphics{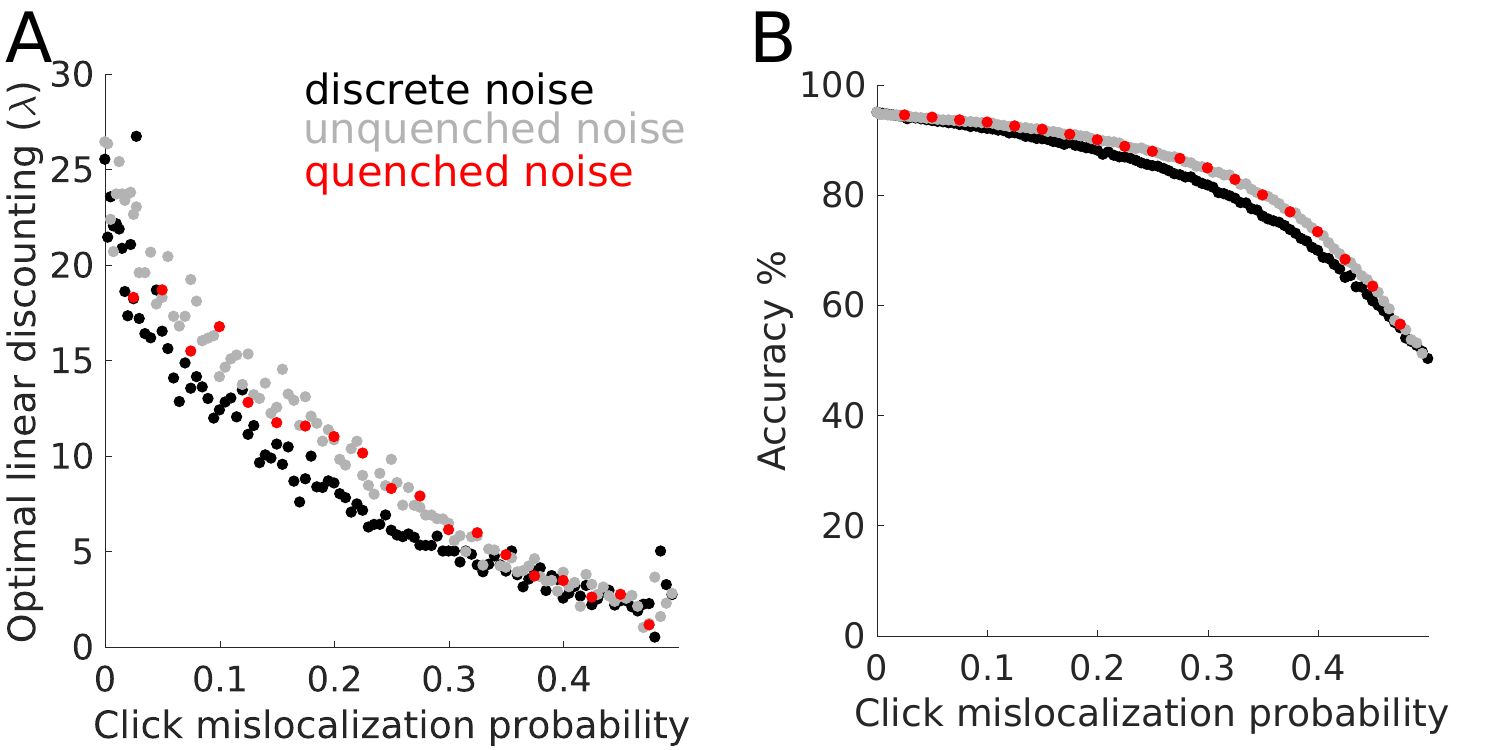}
\caption{\textbf{Three Interpretations of Gaussian noise} Numerically optimized discounting rates for different noise amplitudes. (Black dots) Discrete noise, same points as Figure 2. (Grey dots) unquenched Gaussian noise of the form in equation \eqref{gaussian_noise}. The unqueched fluctuations favor a larger discounting rate. }
\label{unquenched_noise}
\end{figure}

Figure \ref{unquenched_noise} shows the results of the comparison. The discrete noise has a slight decrease in accuracy, and a slightly smaller discounting rate. The difference is due to clicks that weakly change sign. The discrete noise doesn't distinguish between small and large amplitude clicks, where the quenched Gaussian noise does. Importantly, in the noise regime we expect the rats, there is no difference between these interpretations of sensory noise. \newline

\noindent{\textbf{Unquenched Gaussian noise in the quantitative model}}\newline
Gaussian noise subjects the clicks to large amplitude fluctuations in how they are perceived. Our quantitative model handles these fluctuations slightly differently from the normative theory outlined in the section above. First, observe that in the optimal inference theory, the evidence reliability term quenches large amplitude fluctuations. Following the derivation in the section above, $\kappa\left(r_1, r_2,P_r,P_l\right)$ is bounded between $\pm \kappa\left(r_1, r_2\right)$, so the evidence added to the accumulation variable after each click is bounded (``quenched'') and not subjected to large amplitude fluctuations.

Second, we asked whether the presence of large amplitude fluctuations of click amplitudes if they are not quenched, would cause a linear approximation to favor a stronger evidence discounting in order to damper the fluctuations. Specifically, we asked whether an evidence discounting agent with unquenched Gaussian noise: 
\begin{eqnarray}\label{gaussian_noise}
da = \left(\delta_{R,t} - \delta_{L,t}\right)\mathcal{N}(1,\sigma^2) -\lambda a dt,
\end{eqnarray} would maximize accuracy with a larger $\lambda$ than the same click mislocalization strength implemented as quenched noise in the normative theory. Quenched noise as properly implemented in the normative theory would look like:
\begin{eqnarray}\label{quenced_gaussian_noise}
da = \left(\delta_{R,t} - \delta_{L,t}\right)\kappa\left(r_1,r_2,\mathcal{N}(1,\sigma^2)\right) -\lambda a dt.
\end{eqnarray}
Figure \ref{unquenched_noise} shows a comparison between quenched and unquenched Gaussian noise. We find no difference between these interpretations. In panel B, the accuracy of the unquenched Gaussian noise is from the best linear discounting agent, because we do not have a normative theory for unquenched noise (precisely what the simulation was asking to compare). \newline

\noindent{\textbf{Click reliability with missed clicks}}\newline
An alternative form of sensory noise might parameterize the probability that a subject just fails to hear a click at all. Using this framework, we show that missed clicks doesn't change the click reliability function. Assume a click that is generated is not detected at all with probability $m$. Then, the click reliability of a click on the right can be computed as:
\begin{eqnarray}
\kappa(r_1,r_2,m) = \log\left( \frac{r_1(1-m)r_2m + r_1(1-m)(1-r_2)}{r_2(1-m)r_1m + r_2(1-m)(1-r_1)} \right)
\end{eqnarray} We can interpret this expression as the probability of having a click be generated on one side and not missed and a click generated on the other side and missed, or the probability of a click being generated on one side and not being missed and no click is generated on the other side. Given that $\Delta t << 1$, we can remove second order terms in $\Delta t$:
\begin{eqnarray}
\kappa(r_1,r_2,m) = \log\left( \frac{r_1(1-m)}{r_2(1-m)} \right) = \log\left( \frac{r_1}{r_2} \right)
\end{eqnarray} We find the click reliability is independent of the probability of missing a click, $m$.  \newpage

\noindent{\textbf{A general argument for click mislocalization}}\newline
In the previous sections we demonstrated that in the case of missed clicks, or gaussian clicks, mislocalization is necessary for decreasing click reliability. Here we provide a general argument for why that is true under any form of sensory noise. The auditory evidence takes on two possible values $S=\left\lbrace+1,-1\right\rbrace$. Let $y$ be the value of each auditory stimuli after being noisily encoded by the sensory transduction process ($y=f(S)$). If $f()$ maps left and right clicks separately into non-overlapping distributions of click amplitudes, then an ideal observer can bin $y$ into groups $y < 0$ and $y >0$, and perfectly recover the original signal $S$. If $f()$ maps left and right clicks into overlapping distributions, then an observer cannot bin $y$ to perfectly recover the original signal. If the observer uses the same binary binning scheme as before, then the error rate in the recovered signal will be equal to the mislocalization rate. Notice that an observer with perfect knowledge of the distribution of $f()$ can do slightly better by using a different binning scheme. If the observer recognizes that clicks in the domain where the left and right distributions overlap are less trustworthy, then the observer can use multiple bins to discount specifically those clicks near 0. The Gaussian reliability function above $\kappa (r_1,r_2, P_r, P_l)$ can be considered an observer with an infinite number of bins. As seen in figure \ref{unquenched_noise}, this strategy slightly improves accuracy above the two-binning strategy. We thus conclude that click mislocalization is the source of decreasing click reliability from sensory noise. \newline

\clearpage
\noindent{\Large\textbf{Psychophysical Reverse Correlation details}}\newline
Here we present two control analyses on our reverse correlation method. First, we show that our method is not biased by the presence of a lapse rate, unlike logistic regression. Second, we rule out degenerate strategies like deciding based on only the last click.  


\begin{figure}[h]
\centering
\includegraphics[width=.75\textwidth]{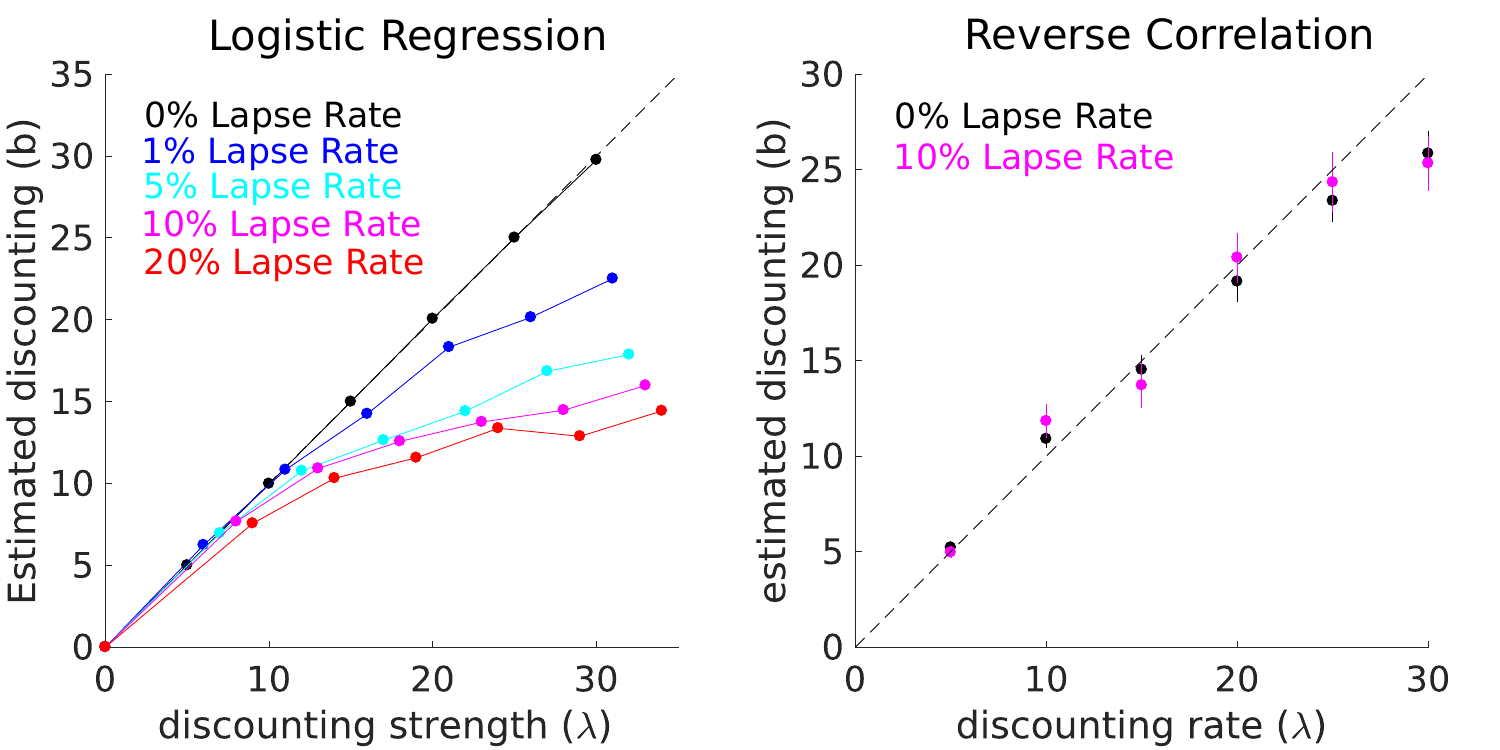}
\caption{\textbf{Reverse Correlation timescales are unaffected by lapse rates}. Lapse rates are defined as the percentage of trials where the subject makes a random response. (A) Logistic regression is strongly biased by the presence of a lapse rate. (B) Psychophysical reverse correlation methods used in this study are not biased. }
\end{figure}\newpage

\clearpage
\noindent {\Large\textbf{Are the rats really integrating? Ruling out last click strategies}}\newline
One possible concern is that the rats might be relying on degenerate strategies like choosing based on the last click they heard. Or that the rat's integration timescale is so short, that their behavior shouldn't really be considering integration. Figure 19A shows a quasi-fixed point analysis of the optimal accumulation equation given a noise level. Assuming the environment stays in one state for a long time, we then replace the evidence term with the expected rate of clicks, and solve for the steady state accumulation value. We can see that for all noise levels, the fixed point lies above 1 click, so the optimal behavior necessarily involves integrating clicks. For the average rat noise level, we see integration of about 5 clicks. 

Figure 19B shows the recovered discounting rate from the reverse correlation method against a simulated discounting agents, similar to Figure 3. Here, we include much stronger discounting agents, and find the recovered discounting rate asymptotes at just under 36 Hz, which is the expected total click rate($r_1 -  r_2 \approx 36$). The last click strategy could be considered a discounting agent with an infinite discounting rate, and would be recovered in our analysis as a discounting rate of about 36. We find our rats are well away from this limit. Thus we confidently rule out a last click strategy. \newline

\begin{figure}[h]
\centering
\includegraphics[width=.7\textwidth]{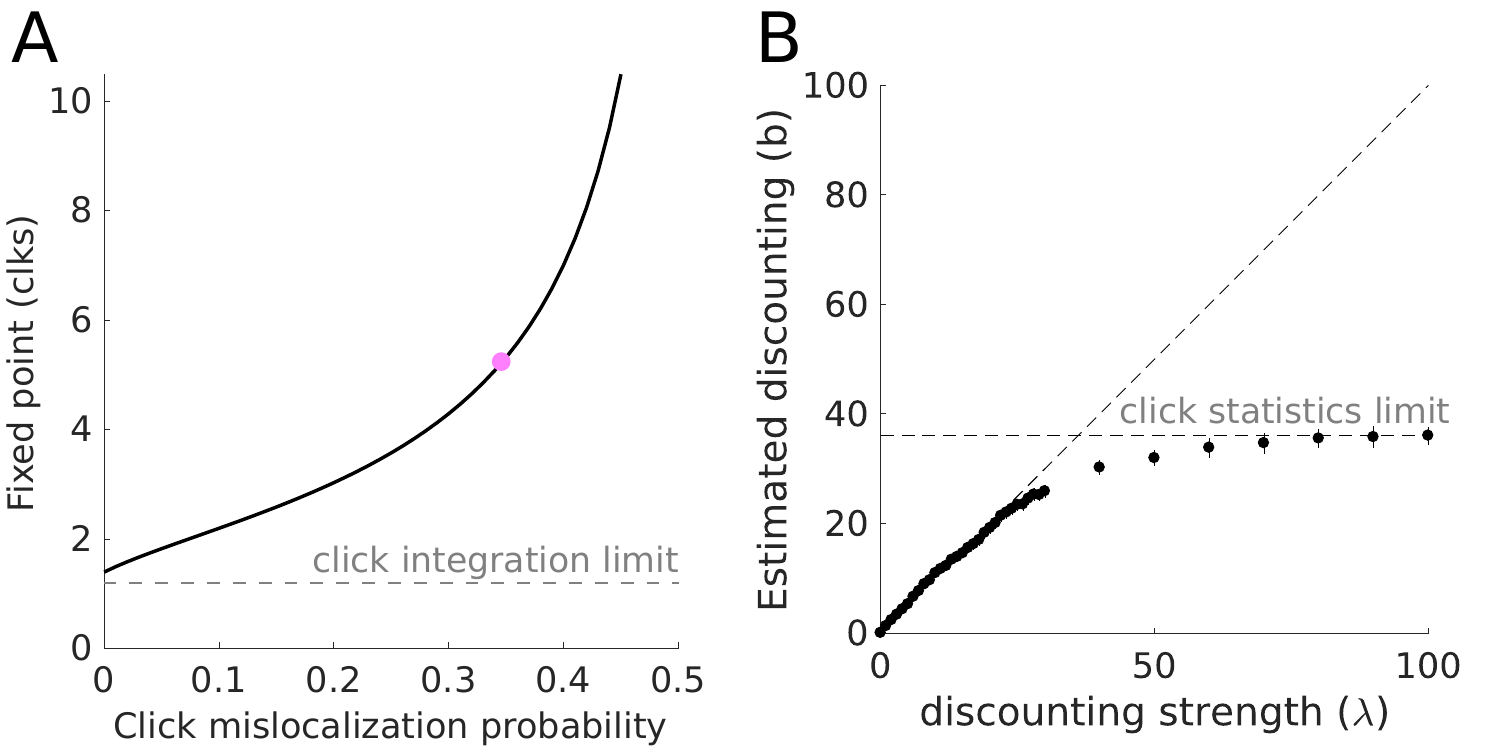}
\caption{\textbf{Ruling out last click strategies.} (A) Quasi fixed points derived from the expected click rate and evidence discounting functions, assuming a fixed environmental state. For all noise levels, the fixed point is greater than 1 click. (B) Integration timescales measured from reverse correlation curves. At large discounting rates, the timescale saturates reflecting the timescale of click generation. }
\end{figure}\newpage

\clearpage
\noindent{\Large\textbf{Model details}}\newline

\begin{figure}[h]
\centering
\includegraphics[width=\textwidth]{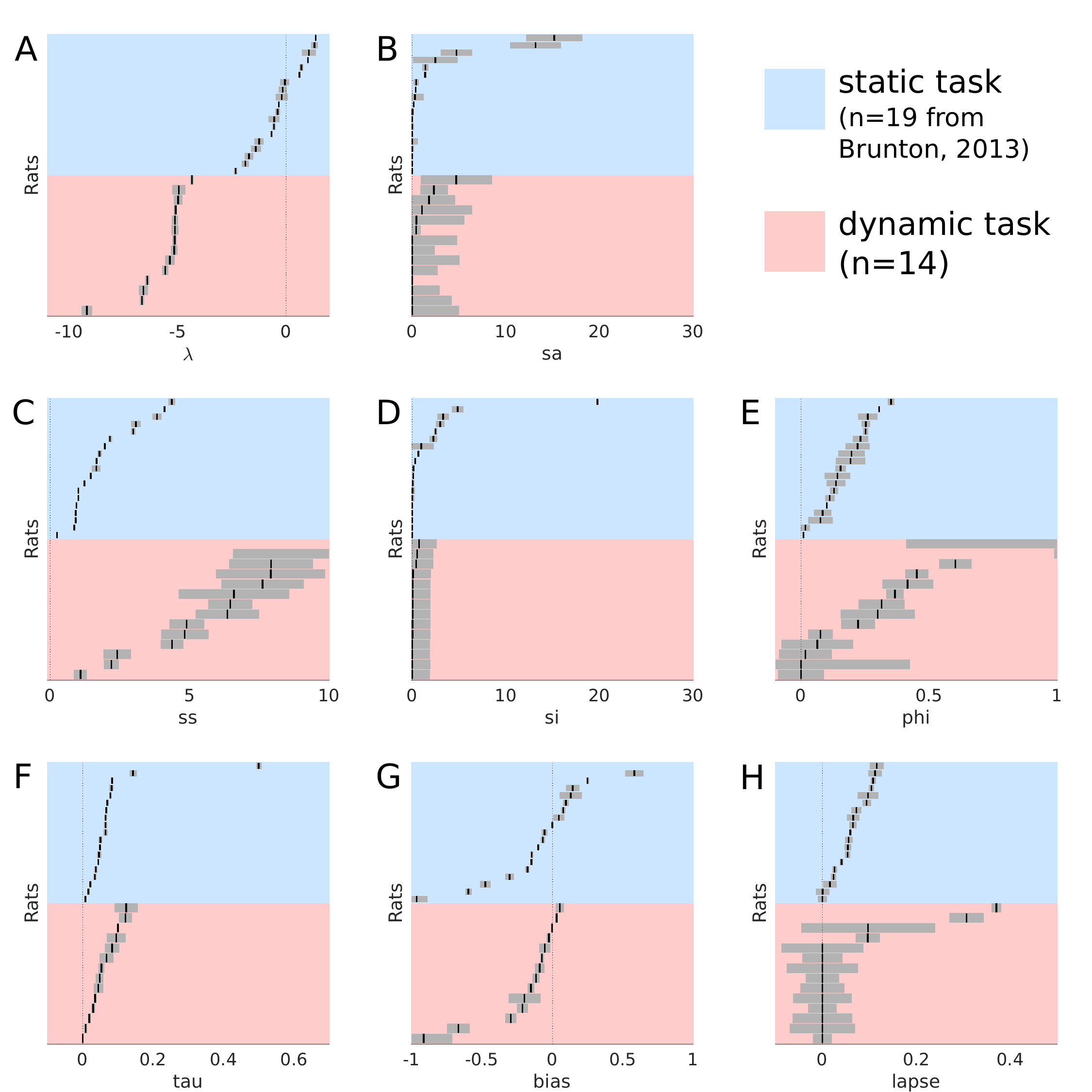}
\caption{\textbf{Best fitting model parameters on static and dynamic tasks.} The best fitting parameters and their standard errors are shown for each rat in the current study, compared to each rat from \citealt{brunton2013}. Each parameter plot has the rats sorted independently by parameter value, rows across panels do not indicate the same rat.}
\end{figure}
\newpage

\begin{sidewaystable}
\small
\centering
{\renewcommand{\arraystretch}{1.2}
\begin{tabular}{|l|l|l|l|l|l|l|l|l||l|}
\hline
rat	   	&  $\lambda$		& $\sigma_s^2$	& $\sigma_a^2$ 	& $\sigma_i^2$ 	& $\phi$ 		& $\tau_{\phi}$ & Bias 			& Lapse 		&  $\lambda_{opt}$	\\
\hline
H033 	& $-5.09 \pm 0.07$& $1.79 \pm 2.76$	& $6.45 \pm 0.79$	& $0.060 \pm 1.87$	& $0.45 \pm 0.045$	& $0.053 \pm 0.0073$	& $-0.025 \pm 0.014$& $1.1e-6 \pm 0.042$	&	$-5.03 \pm 0.014$\\
H037 	& $-6.39 \pm 0.092$& $0.41 \pm 0.44$	& $2.19 \pm 0.26$   & $0.0076 \pm 1.87$	& $0.37 \pm 0.032$	& $0.12 \pm 0.018$	& $0.030 \pm 0.0072$ &	$3.6e-9 \pm 0.020$ 	&	$-5.92 \pm 0.0079$\\
H038 	& $-4.33 \pm 0.063$& $2e-4 \pm 4.99$	& $11.6 \pm 0.99$	& $0.71 \pm 1.86$	& $1.20 \pm 0.79$	& $1.2e-4 \pm 1.7e-4$			& $-0.21 \pm 0.039$& $0.097 \pm 0.025$	&   $-5.13 \pm 0.091$\\
H039 	& $-4.94 \pm 0.30$	&$0.44 \pm 5.11$& $11.52 \pm 4.96$	& $0.052 \pm 1.87$	& $5.3e-4 \pm 0.42$	& $0.0083 \pm 0.0042$ 	& $-0.91 \pm 0.20$	& $0.097 \pm 0.14$	&	$-4.56 \pm 0.13$	\\
H040 	& $-6.64 \pm 0.099$	& $2.30 \pm 1.45$	& $4.36 \pm 0.40$	& $0.072 \pm 1.87$	& $0.075 \pm 0.047$	& $0.035 \pm 0.0034$ 	& $-0.073 \pm 0.012$	& $1.9e-7 \pm 0.029$	&	$-5.53 \pm 0.079$	\\
H043 	& $-4.98 \pm 0.19$	& $1.04 \pm 5.3$	& $7.61 \pm 1.47$	& $0.052 \pm 1.87$	& $1.02 \pm 0.029$	& $0.12 \pm 0.032$ 	& $-0.20 \pm 0.11$	& $0.31 \pm 0.036$	&	$-5.94 \pm 0.0022$ \\
H045 	& $-5.11\pm0.13$	& $0.0029 \pm 2.37$	& $1.09 \pm 0.23$	& $0.028 \pm 1.87$	& $4.8e-4 \pm 0.089$	& $0.048 \pm 0.010$	& $-0.67 \pm 0.079$	& $0.37 \pm 0.0098$	&	$-5.27 \pm 0.18$	\\
H058 	& $-5.15 \pm 0.15$  & $3.59e-5 \pm 2.92$& $4.82 \pm 0.84$ & $6.64e-5 \pm 1.87$ & $0.42 \pm 0.098$ & $0.10 \pm 0$ & $-0.15 \pm 0.023$ & $1.64e-7 \pm 0.063$ & $-4.87 \pm 0.42$ \\
H061 	&  $-5.12 \pm 0.15$	& $4.69\pm 3.78$&$6.59\pm 1.97$	&$0.038\pm 1.87$	&$0.31 \pm 0.089$	&$0.067 \pm 0.019$ 	&$0.053 \pm 0.026$	& $1e-6 \pm 0.087$	& $-4.75 \pm 0.0026$	\\
H065 	& $-6.57 \pm 0.20$	&$1.53e-5 \pm 4.99$	& $7.91 \pm 1.49$	& $0.0065 \pm 1.87$	& $0.064 \pm 0.14 $	& $0.019 \pm 0.0040$ 	& $-0.089 \pm 0.033$	& $4.09e-8 \pm 0.069$	&	$-5.37 \pm 0.0054$	\\
H066 	& $-5.13 \pm 0.089$	& $5.9e-5 \pm 2.68$	& $4.89 \pm 0.61$	& $0.52 \pm 1.71$	& $0.60 \pm 0.063$	& $0.084 \pm 0.019$ 	& $-0.12 \pm 0.022$	& $3.8e-7 \pm 0.035$	& $-5.58 \pm 0.27$	\\
H067 	& $-9.18 \pm 0.23$	& $3.9e-5 \pm 0$ & $2.40 \pm 0.49$	& $0.40 \pm 1.84$	& $0.22 \pm 0.065$	& $0.095 \pm 0.026$ 	& $-0.0012 \pm 0.077$	& $3.3e-7 \pm 0.047$	& $-6.05 \pm 0.081$	\\
H083 	& $-5.35 \pm 0.22$	& $0.0034 \pm 4.75$	& $7.90 \pm 1.95$	& $0.028 \pm 1.87$	& $0.30 \pm 0.14$	&$0.045 \pm 0.013$ 	& $-0.053 \pm 0.038$	& $9.5e-7 \pm 0.075$	&	$-4.87 \pm 0.025$ \\
H084 	& $-5.56 \pm 0.014$	& $3.5e-5 \pm 4.21$	& $6.35 \pm 1.12$	& $1.02e-4 \pm 1.91$	& $0.017 \pm 0.103$	& $0.029 \pm 0.0054$ 	& $-0.29 \pm 0.037$&$2.3e-7 \pm 0.062$	& $-5.17 \pm 0.12$	\\
\hline
\end{tabular}}
\caption{\doublespacing\textbf{Maximum likelihood parameters and the standard error for each parameter.} }
\end{sidewaystable}
\clearpage

\begin{figure}
\centering
\includegraphics[width=\textwidth]{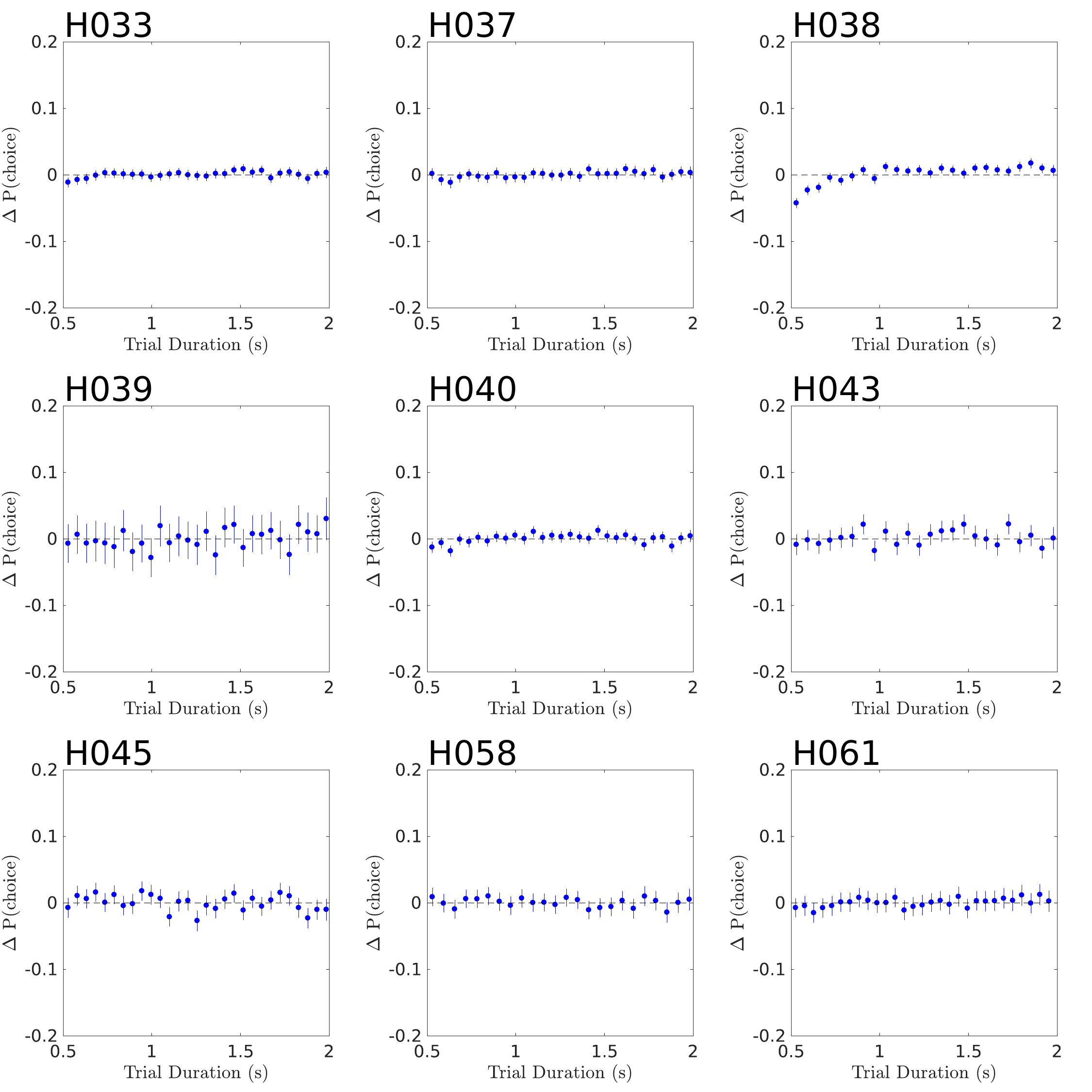}
\caption{\textbf{Model Residual error against time} The model fits short and long trials equally well.}
\end{figure}
\newpage

\begin{figure}
\centering
\includegraphics[width=\textwidth]{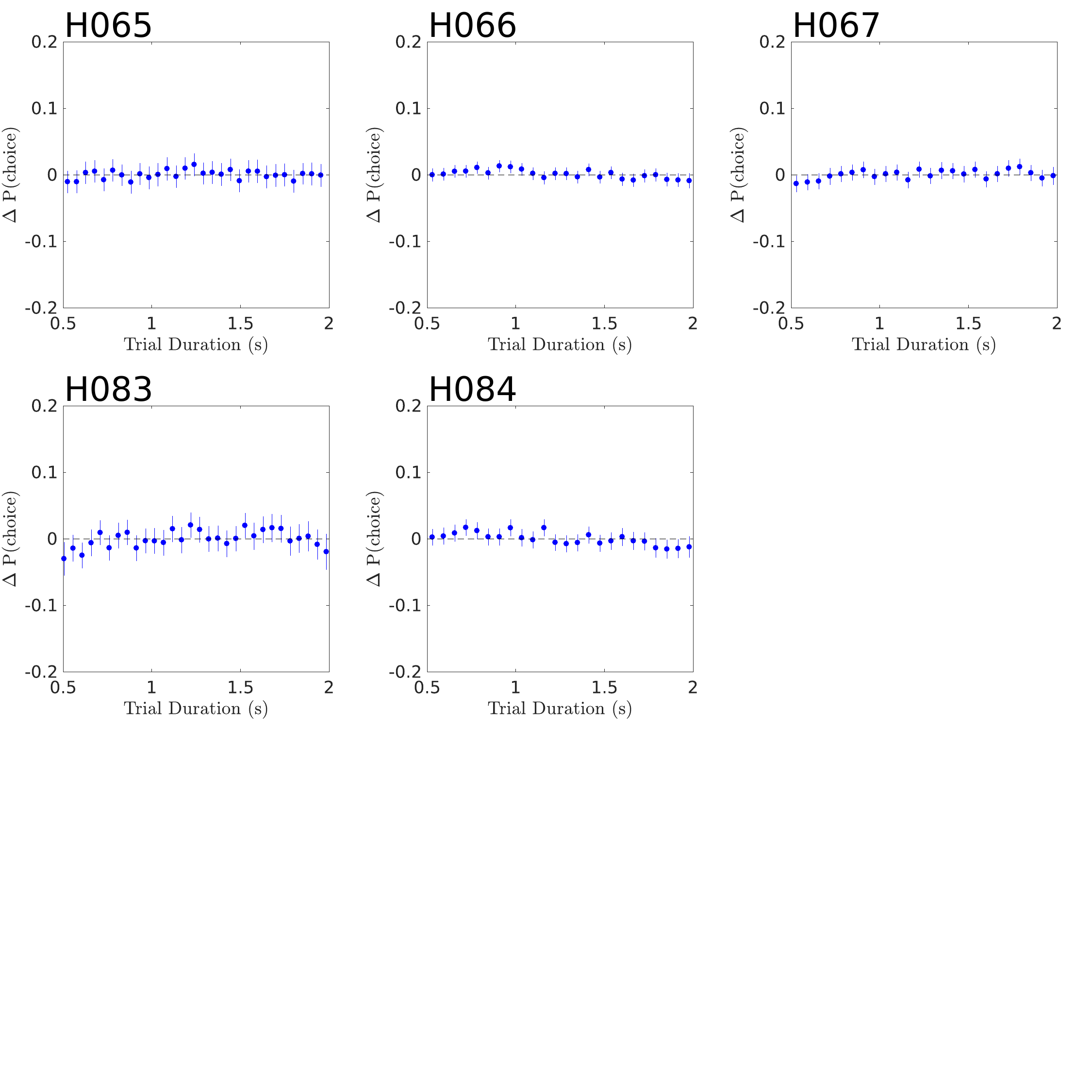}
\caption{\textbf{Model Residual error against time} The model fits short and long trials equally well.}
\end{figure}
\clearpage

\begin{figure}
\centering
\includegraphics[width=\textwidth]{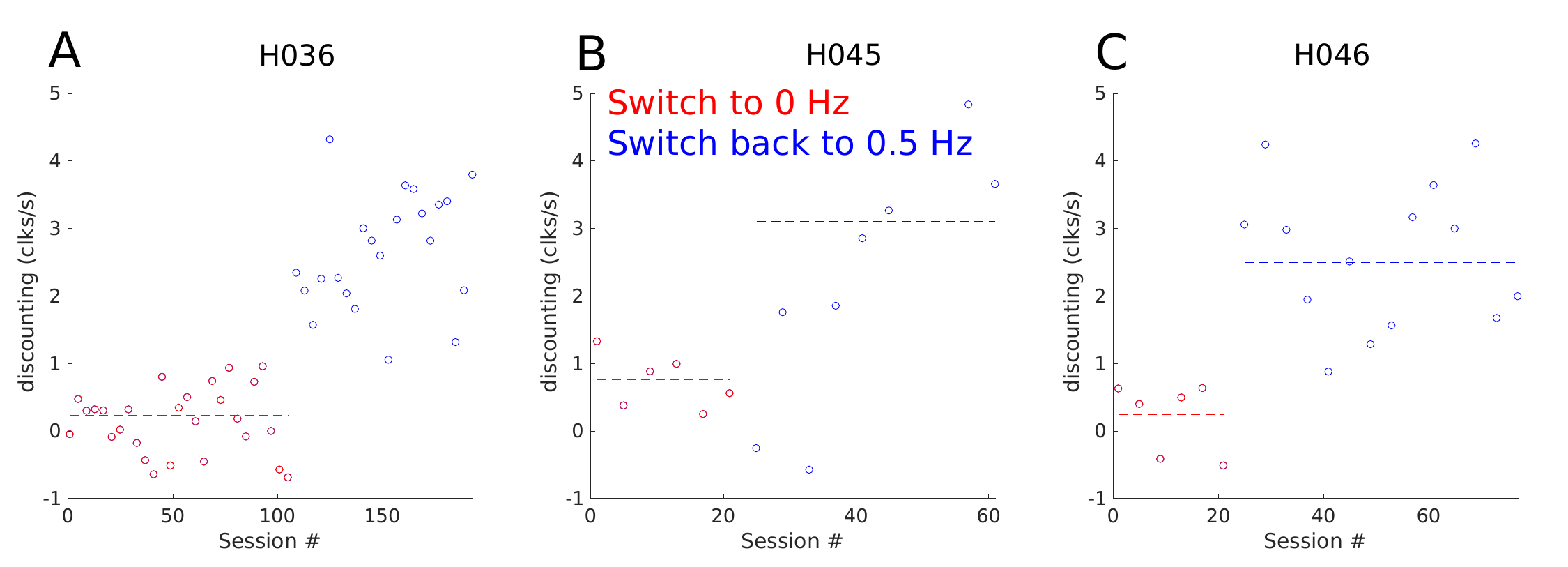}
\caption{\textbf{Rats adjust their integration timescales quickly to new environments} Evidence discounting rates estimated in blocks of 4 sessions for each rat in figure 6D. Session 1 is the first session in the 0Hz environment. Each rat is then moved back to 0.5 Hz. Dashed lines show the evidence discounting rates estimated over all sessions of the same hazard rate. Variability across blocks of session is due to low trial count. }
\end{figure}
\newpage

\clearpage
{\footnotesize \bibliography{references.bib}}
\bibliographystyle{apalike}
\end{document}